\documentclass[aps,prb,twocolumn,showpacs,superscriptaddress,final,floatfix]{revtex4-1}  % for review and submission
\usepackage[pdftex]{graphicx}  % needed for figures
\usepackage[update]{epstopdf}
\usepackage{dcolumn}   % needed for some tables
\usepackage{bm}        % for math
\usepackage{amssymb}   % for math
\usepackage{amsmath}   % for math
\usepackage{color}
\usepackage{array}
\usepackage{dcolumn}
\graphicspath {{./FIG/}}
\makeatletter
\def\input@path{{./FIG/}}
\makeatother
\def\d{\mathrm{d}\hspace*{-0.1ex}}
\newcolumntype{t}[1]{D{.}{.}{#1}}
\def\cdash{\multicolumn{1}{c}{\textemdash}}

\begin{document}

\title{Domain-wall excitations in the two-dimensional Ising spin glass}

\author{Hamid Khoshbakht}
\email{hamidkhoshbakht@gmail.com}
\affiliation{Applied Mathematics Research Centre, Coventry University,
  Coventry, CV1~5FB, United Kingdom}
\affiliation{Institut f\"ur Physik, Johannes Gutenberg-Universit\"at Mainz,
  Staudinger Weg 7, D-55099 Mainz, Germany}
\author{Martin Weigel}
\email{martin.weigel@complexity-coventry.org}
\affiliation{Applied Mathematics Research Centre, Coventry University,
  Coventry, CV1~5FB, United Kingdom}

\date{\today}

\begin{abstract}

  The Ising spin glass in two dimensions exhibits rich behavior with subtle
  differences in the scaling for different coupling distributions. We use recently
  developed mappings to graph-theoretic problems together with highly efficient
  implementations of combinatorial optimization algorithms to determine exact ground
  states for systems on square lattices with up to $10\,000\times 10\,000$
  spins. While these mappings only work for planar graphs, for example for systems
  with periodic boundary conditions in at most one direction, we suggest here an
  iterative windowing technique that allows one to determine ground states for fully
  periodic samples up to sizes similar to those for the open-periodic case. Based on
  these techniques, a large number of disorder samples are used together with a
  careful finite-size scaling analysis to determine the stiffness exponents and
  domain-wall fractal dimensions with unprecedented accuracy, our best estimates
  being $\theta = -0.2793(3)$ and $d_\mathrm{f} = 1.273\,19(9)$ for Gaussian
  couplings. For bimodal disorder, a new uniform sampling algorithm allows us to
  study the domain-wall fractal dimension, finding $d_\mathrm{f} = 1.279(2)$.
  Additionally, we also investigate the distributions of ground-state energies, of
  domain-wall energies, and domain-wall lengths.

\end{abstract}

\maketitle

\section{Introduction}

The problem of an adequate description and understanding of the behavior of spin
systems with strong disorder has been studied for around forty years by a large
number of scientists in statistical and condensed matter physics as well as,
increasingly, researchers in adjacent fields such as computer science and mathematics
\cite{kawashima:03a}. It is a hard problem in that many of the well-developed tools
of the theory of critical phenomena, such as the renormalization group, fail to
satisfactorily describe all important aspects of these models, and in that the
standard techniques of numerical simulations are faced with diminishing efficiency in
view of exploding relaxation times and the massive computational demand of the
average over quenched disorder. But it is also a good and fruitful problem in that
the questions it poses are deeply rooted in the foundations of statistical mechanics
\cite{binder:86a} and the simplicity of the models has led to applications ranging
from the physics of structural glasses to error correcting codes and neural networks
\cite{nishimori:book}.

While even fundamental questions such as the values of the lower and upper critical
dimensions of such models are still under active debate \cite{viet:09,fernandez:09a,
  viet:10,sharma:11,beyer:12}, there is consensus that a spin-glass phase appears at
non-zero temperatures for short-ranged systems of Ising spins in at least three
dimensions, but no spin-glass order occurs beyond ground states in two-dimensional
(2D) systems \cite{hasenbusch:08,ohzeki:09,fernandez:16}. While such 2D geometries
might hence appear less useful for modeling experimentally realized spin-glass
phases, the physics of these systems is in fact rather interesting in its own
right. One intriguing aspect is that for sufficiently asymmetric coupling
distributions a long-range ferromagnetic phase can exist at non-zero temperatures,
and it is found that the phase boundary at low temperatures shows re-entrance or
inverse melting, that is, on further cooling a system in the ferromagnetic phase,
order is lost in favor of a paramagnetic state \cite{toldin:08,thomas:11a}. Another
facet is the question of universality regarding the distribution of exchange
couplings: at zero temperature, the bimodal model has extensive ground-state
degeneracies leading to behavior rather different from the case of continuous
coupling distributions \cite{hartmann:01a}. The resulting entropy of volatile spin
clusters was long believed to lead to power-law correlations at zero temperature, but
there is now evidence of true long-range spin-glass order \cite{jorg:06,roma:10}. The
behavior of this model at low temperatures is determined by a delicate interplay of
the distinct fixed points of the universality classes of discrete and continuous
coupling distributions, respectively
\cite{thomas:11,toldin:11a,joerg:12,jinuntuya:12}, and there is still no complete
consensus about universality at finite temperatures \cite{fernandez:16,lundow:16}. It
is the subtle role played by entropic fluctuations which makes this model relevant to
the finite-temperature transitions observed in three dimensions \cite{thomas:11}.

Apart from such theoretical considerations, interest in the 2D models has been fueled
by the relative ease in numerical tractability as compared to higher-dimensional
systems. This goes beyond the general advantage of systems in low dimensions of
providing larger linear system sizes at the same number of sites: 2D systems in zero
external field are an exception to the {\em NP\/} hardness of ground-state problems
found in systems of higher dimensions \cite{barahona:82}. Ground states on planar
graphs can be determined in polynomial time from the mapping to a minimum-weight
perfect matching problem \cite{bieche:80a}. This allows to treat significantly larger
lattice sizes than those accessible to simulation methods. The restriction to planar
graphs, and hence periodic boundary conditions in at most one direction, has been
rather inconvenient for certain types of studies \cite{hartmann:02} and, in general,
leads to relatively larger finite-size corrections. Polynomial-time algorithms also
exist for the more general problem of determining the partition function
\cite{blackman:91,saul:93,galluccio:00}. These methods, based on the evaluation of
Pfaffians, have the advantage of allowing for periodic boundary conditions, but they
are technically more demanding than the ground-state computations and thus restricted
to smaller system sizes. Only recent advances have allowed to extend these approaches
to system sizes $L\gtrsim 100$. \cite{thomas:09} In parallel, exact sampling
techniques for Ising spin glasses at non-zero temperatures based on the application
of ``coupling-from-the-past'' \cite{propp:96} or sampling of dimer coverings
\cite{wilson:97} have recently been suggested, that are either restricted to or only
efficient in 2D \cite{chanal:08,thomas:09}.

A wide range of aspects of 2D spin glasses has been found to be consistent with
droplet theory \cite{mcmillan:84,bray:87a,fisher:88}. Droplet and domain-wall
excitations can be directly inserted in zero-temperature configurations. Domain-wall
energies are found to scale as a power law
$E_\mathrm{def}\sim L^{\theta_\mathrm{DW}}$ for the Gaussian model with
$\theta_\mathrm{DW}\approx -0.28$.  \cite{hartmann:01a} Roughly consistent values are
found for the scaling of droplet energies if scaling corrections are taken into
account \cite{hartmann:03a}. No power-law scaling of domain-wall energies is found
for bimodal couplings \cite{hartmann:01a}, but droplets in this model show
$\theta\approx-0.29$, possibly compatible with the Gaussian case
\cite{hartmann:08}. As the spin-glass phase is confined to zero temperature for 2D
models, ground-state calculations give direct access to the {\em critical\/} behavior
of the spin-glass transition. In this case, the correlation length exponent is
expected to follow from $\nu = -1/\theta$. \cite{bray:87a} As $\eta = 0$ at least for
the Gaussian model \cite{fernandez:16}, this is the only relevant critical exponent
(but see Ref.~\onlinecite{hartmann:08} for the bimodal case).  Domain walls and
droplet interfaces are found to be fractal curves with dimension $d_\mathrm{f} < 2$,
i.e., not space filling \cite{bray:87}. At least for the Gaussian case, these fractal
curves appear to be compatible, under certain conditions, with a description in terms
of stochastic Loewner evolution \cite{amoruso:06a,bernard:07,khoshbakht:12}. Such
consistence together with further assumptions would suggest a relation between
stiffness exponent and fractal dimension, $d_\mathrm{f} = 1+3/[4(3+\theta)]$.
\cite{amoruso:06a} For the bimodal model, on the other hand, the fractal dimension is
possibly different \cite{melchert:07,weigel:06a,gusman:08}, but calculations are
complicated by sampling problems since the ground-state algorithms do not produce the
degenerate ground states with the correct weights. These subtle differences between
results for different coupling distributions and excitation types call for
high-precision studies to distinguish random from systematic coincidences. Some
previous results for $\theta$ and $d_\mathrm{f}$ in the Gaussian model are collected
in Table \ref{theta_articles}.

Here we combine a formulation of the ground-state problem on planar graphs in terms
of Kasteleyn cities \cite{gregor:07,thomas:07} with a recently suggested efficient
implementation of the Blossom algorithm for minimum-weight perfect matching
\cite{kolmogorov:09}. This allows us to determine ground states for systems of up to
$10\,000 \times 10\,000$ spins on commodity hardware. To extend these results to the
case of periodic boundaries with the smaller scaling corrections expected there, we
introduce a hierarchical optimization procedure using windows, alike to the patchwork
dynamics discussed in Ref.~\onlinecite{thomas:08}, which allows to determine ground
states of fully periodic samples with a constant relative increase in computational
effort as compared to the matching technique for planar samples. To treat the case of
bimodal couplings correctly, we use a new approach based on an exact decomposition of
the ground-state manifold into rigid clusters that are then sampled within a parallel
tempering framework that guarantees uniform sampling of ground states to high
precision.

The rest of this paper is organized as follows. In Sec.~\ref{sec:model} we outline
the matching algorithm based on Kasteleyn cities, introduce the windowing technique
that allows to generalize the method to systems with fully periodic boundaries, and
evaluate the performance of these algorithms. Section \ref{sec:gauss} is devoted to
the system with Gaussian coupling distribution, and we report our results for the
average ground-state and defect energies, the domain-wall fractal dimension as well
as the probability distributions of these quantities for different boundary
conditions. In Sec.~\ref{sec:bimodal} we analyze these quantities for the bimodal
model, introducing a new uniform-sampling technique for the degenerate ground states
in this case that allows us to provide an unbiased estimate of the domain-wall
fractal dimension. Finally, Sec.~\ref{sec:conclusions} discusses the compatibility of
our results with the conjecture $d_\mathrm{f} = 1+3/[4(3+\theta)]$ of
Ref.~\onlinecite{amoruso:06a} and contains our conclusions.

\begin{table}[tb!]
\caption{Previous estimates of the spin-stiffness exponent $\theta$ and the fractal
  dimension $d_\mathrm{f}$ of the 2D Ising spin glass with Gaussian bound distribution.}
\begin{ruledtabular}
\begin{tabular}{lt{4}t{4}c}
  Ref.    &\multicolumn{1}{c}{$\theta$} &  \multicolumn{1}{c}{$d_\mathrm{f}$} & \multicolumn{1}{c}{max. system size} \\ \hline
  \onlinecite{mcmillan:84a} &  -0.281(5)   &  \cdash     & $8\times8$       \\
  \onlinecite{palassini:99} &   -0.285(2)  &  \cdash     & $30\times30$     \\
  \onlinecite{bray:84}      &  -0.294(9)   &  \cdash     & $12\times 12$    \\ 
  \onlinecite{bray:87}      &  -0.29(1)    & 1.26(3)       & $120\times13$   \\
  \onlinecite{rieger:96}    &   -0.281(2)  &  1.34(10)    & $30\times30$  \\
  \onlinecite{hartmann:01a} &  -0.282(2)   &  \cdash     & $480\times480$  \\
%  \onlinecite{hartmann:01a} &  -0.266(2)   &  \cdash     & $480\times480$    \\ 
  \onlinecite{middleton:01} &  \cdash      &  1.25(1)    & $256\times256$    \\ 
  \onlinecite{weigel:06a}   &   -0.284(4)  &  1.273(3) & $256\times256 $    \\
  \onlinecite{bernard:07}   &   \cdash     &  1.28(1)    & $720\times 360$    \\
  \onlinecite{hartmann:02a} &   -0.287(4)  &  \cdash     & $16\times1024$  \\
  \onlinecite{carter:02a}   &  -0.282(3)   &  \cdash     & $12\times384$   \\ 
  \onlinecite{hartmann:04a} &  -0.281(7)  &  \cdash     & $64\times64$\\
  \onlinecite{amoruso:06a}  &  -0.285(5)  & 1.27(1)       & $300\times300 $   \\
  \onlinecite{melchert:07}  &  -0.287(4)  & 1.274(2)   & $320\times320$    \\ 
%  \onlinecite{melchert:09}  &  -0.287(4)  &  1.274(2)      & $512\times512$   \\
\hline
This work           & -0.2793(3)  & 1.27319(9)  & $10\,000\times10\,000$ \\
\end{tabular}
\end{ruledtabular}
\label{theta_articles}
\end{table}

%$$$$$$$$$$$$$$$$$$$$$$$$$$$$$$$$$$$$$$$$$$$$$$$$$$$$$$$$$$$$$$$

\section{Model and algorithms\label{sec:model}}

\begin{figure*}
  \includegraphics[width=0.95\textwidth]{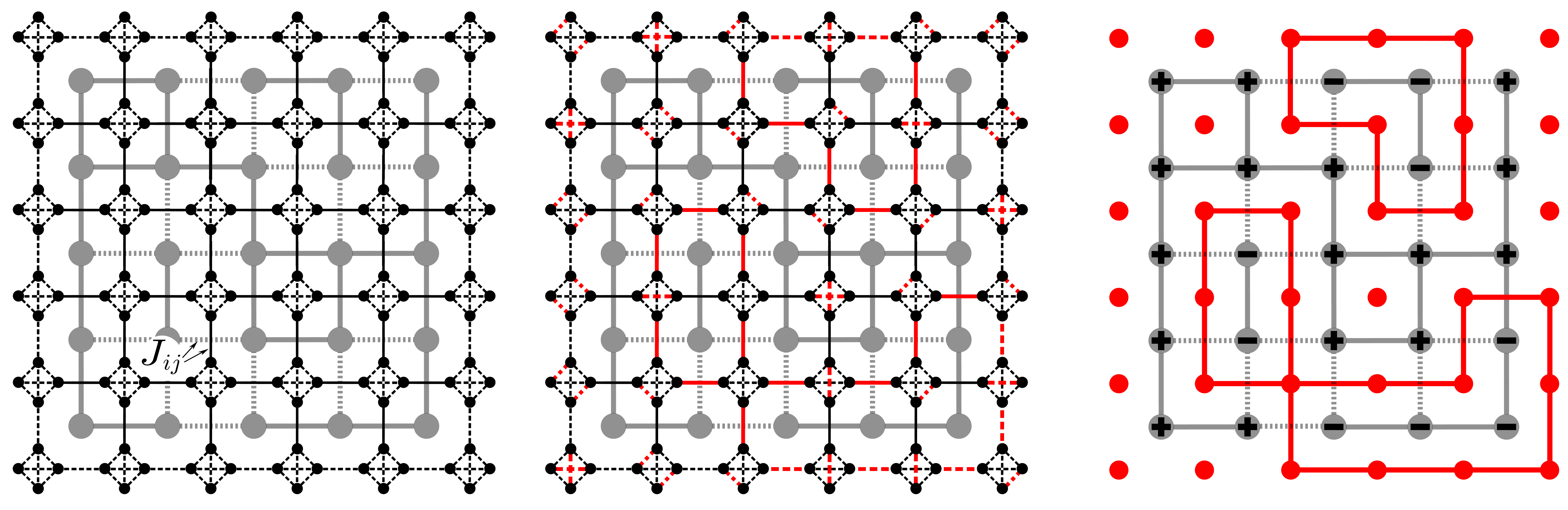}
  \caption{
    (Color online)
    Mapping of the Ising spin-glass ground-state problem to a minimum-weight perfect
    matching. An auxiliary graph is constructed by expanding each plaquette of the
    dual lattice into a complete graph $K_4$ of four nodes (left). Additional rows and
    columns of $K_4$ nodes are added instead of the outer plaquette to make the
    auxiliary graph more regular. Edge weights on the auxiliary graph are $J_{ij}$
    for each bond that crosses a bond $(i,j)$ of the original graph and zero
    otherwise. Then, a minimum-weight perfect matching is determined on the auxiliary
    graph (middle). By contracting the $K_4$ vertices again, the matching reduces to
    a minimum cut on the spin lattice, i.e., a set of closed loops surrounding
    islands of down spins in a sea of up spins or vice versa (right). Dashed bonds on
    the spin lattice correspond to antiferromagnetic couplings $J_{ij} < 0$, solid
    bonds to ferromagnetic ones, $J_{ij} > 0$.
    \label{fig:kasteleyn}
  }
\end{figure*}

\subsection{The model}

We consider the random-exchange, zero-field Ising model with Hamiltonian
\begin{equation}
  {\cal H} = -\sum_{\langle i,j\rangle} J_{ij} s_i s_j.
  \label{eq:hamiltonian}
\end{equation}
Here, $\langle i,j\rangle$ denotes summation over pairs of nearest neighbors. For the
purposes of this study, the underlying lattice is chosen to have square elementary
plaquettes, but the techniques described here are applicable {\em mutatis mutandis\/}
to any regular planar graph (see, for instance,
Refs.~\onlinecite{weigel:06a,melchert:11}). The case of non-planar graphs is discussed
in Sec.~\ref{sec:window} below.

The couplings $J_{ij}$ are quenched random variables. At zero temperature, two
distinct types of behavior are expected, one for discrete and commensurate allowed
coupling values and a second class for distributions with incommensurate or
continuous support \cite{amoruso:03a,jorg:06,thomas:11,toldin:11a,joerg:12}. We
consider one representative of each class, namely the symmetric bimodal ($\pm J$)
distribution,
\begin{equation}
  P(J_{ij}) = \frac{1}{2}\delta(J_{ij}-J)+\frac{1}{2}\delta(J_{ij}+J),
  \label{eq:bimodal}
\end{equation}
for the commensurate class and the symmetric Gaussian,
\begin{equation}
  J_{ij} \sim {\cal N}(0,1),
  \label{eq:gaussian}
\end{equation}
as example of the continuous class of distributions.

\subsection{Matching with Kasteleyn cities}
\label{sec:kasteleyn}

It was initially noted by Toulouse that the model \eqref{eq:hamiltonian} could be
dualized and the trivial up-down symmetry of the states removed by considering the
interactions around an elementary plaquette \cite{toulouse:77a}. Each plaquette with
an odd number of antiferromagnetic bonds is inherently {\em frustrated\/}, such that
in each spin configuration at least one of the elementary interactions around the
plaquette will be unsatisfied. The energy of the ground state of such a system will
hence be elevated above the ground-state energy of a ferromagnet by an amount
proportional to the total weight of such {\em broken\/} bonds. If edges of the {\em
  dual\/} lattice are used to indicate the broken bonds, these link together to form
defect lines on the dual lattice, emanating and ending in frustrated plaquettes
\cite{bieche:80a}. The search for a ground state is thus (for a planar lattice)
equivalent to the determination of a {\em minimum-weight perfect matching\/} (MWPM)
on the complete graph of frustrated plaquettes, where the edge weights correspond to
the shortest paths (on the dual lattice) between each pair of frustrated
plaquettes. For details see, e.g., Refs.~\onlinecite{bieche:80a,weigel:06a}. As MWPM
is a polynomial problem which is solved efficiently using the so-called blossom
algorithm \cite{edmonds:65a}, it was first noted by Bieche {\em et al.\/}
\cite{bieche:80a} that this allows to calculate exact ground states for relatively
large systems.

\begin{figure}
  \begin{center}
    \includegraphics[width=2.8 in, height=3.0 in, angle=0]{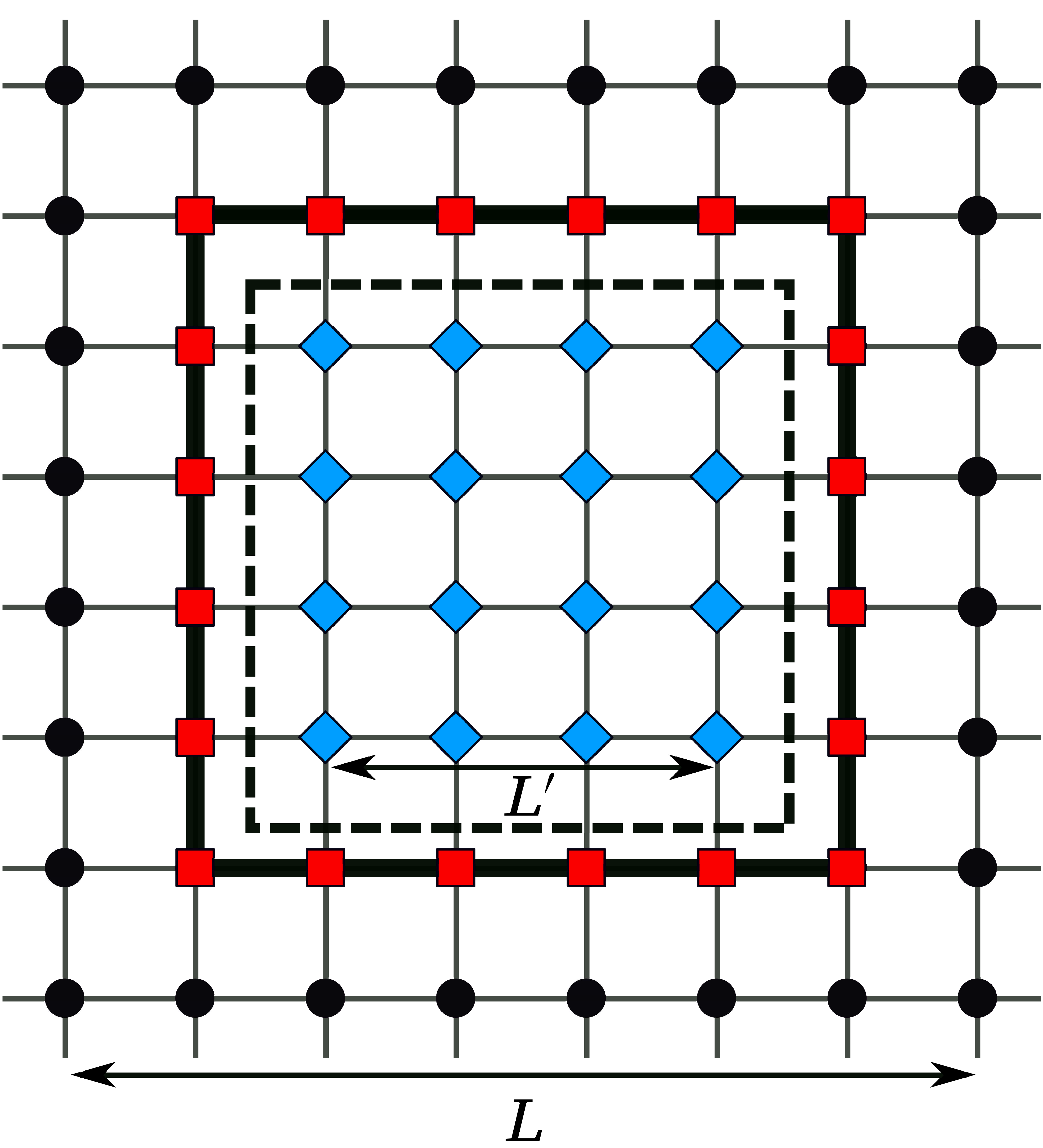}
    \caption{%
      (Color online) Schematic representation of the windowing technique to determine
      ground states for toroidal systems. The  dashed square shows 
      the window and the blue diamonds represent the sites whose spins will be updated next by 
      the windowing technique, as they are contained  within the current
      window. Red squares indicate sites whose spins are fixed in their current orientation with
      strong bonds, indicated by the thick black lines. As a result, the MWPM problem
      will be solved for the system of red and blue spins with using free
      boundary conditions.
    }
    \label{fig:windowing_technique}
  \end{center}
\end {figure}

In practice, however, the outlined mapping has certain disadvantages. The weighted
distance between each pair of frustrated plaquettes needs to be determined before the
matching can proceed. Since each plaquette could be matched up with any other, a
solution is sought for the {\em complete\/} graph of frustrated plaquettes. The
average number of such plaquettes is $F=\alpha N$, where $N$ is the number of spins
and $\alpha$ is a disorder-dependent constant that equals $\alpha = 1/2$ for the
symmetric distributions considered here. The number of edges, however, is $F(F-1)/2$,
increasing quadratically in the system volume. The original implementation of the
blossom algorithm has complexity $O(V^2E)$, where $V$ is the number of vertices in
the auxiliary graph and $E$ the number of edges \cite{edmonds:65a}. For the present
problem, this corresponds to $O(L^8)$ scaling. Memory requirements are
$O(L^4)$. While a number of algorithms with improved worst-case complexity have been
proposed, not all of them are fast and hence useful in practice. We use here the
currently fastest publicly available algorithm due to Kolmogorov
\cite{kolmogorov:09}.  As it is unlikely that edges with a very large weight are part
of the minimum-weight matching, in practice only edges up to a certain weight are
retained \cite{bieche:80a}. One has to proceed carefully here, however, to ensure
high success rates also for larger system sizes. Strictly speaking, the resulting
algorithm is merely quasi-exact.

A polynomial-time solution to the Ising spin-glass ground state problem on planar
graphs based on a somewhat different mapping was proposed in
Refs.~\onlinecite{thomas:07,gregor:07}. This is a rather direct implementation of the
interpretation of the Ising ground-state search as a maximum/minimum-cut
problem. Splitting the Hamiltonian \eqref{eq:hamiltonian} into three terms as
follows,
\begin{equation}
  -{\cal H} = W^++W^--W^\pm = K-2W^\pm,
\end{equation}
where $K = \sum_{\langle ij\rangle}J_{ij}$ and
\begin{equation}
  \begin{split}
    W^+ &= \sum_{\stackrel{\langle ij\rangle}{s_i=s_j=+1}}J_{ij},\\
    W^- &= \sum_{\stackrel{\langle ij\rangle}{s_i=s_j=-1}}J_{ij},\\
    W^\pm &= \sum_{\stackrel{\langle ij\rangle}{s_i\ne s_j}}J_{ij},
  \end{split}
\end{equation}
it is clear that the energy is minimized for a configuration that minimizes $W^{\pm}$
which is the weight of the {\em cut\/} or, in more physical terms the interface,
separating up-spins from down-spins. Note that the interface can consist of more than
one connected component. As it turns out, such cuts can be related one-to-one to
perfect matchings in an auxiliary graph. To see this, consider the example shown in
Fig.~\ref{fig:kasteleyn}. The right panel shows a configuration of up and down spins
on a patch of the square lattice with free boundaries together with the corresponding
cut of anti-aligned neighboring spins. The cut forms a set of closed loops on the
dual lattice (red lines). To represent it as a matching, consider the auxiliary graph
shown on the left of Fig.~\ref{fig:kasteleyn} that replaces each plaquette of the
original lattice (i.e., each node of the dual lattice) by a complete graph of four
nodes, a ``Kasteleyn city''. To create a regular lattice graph, the single outer
plaquette of the dual graph is replaced by $4L$ individual plaquettes surrounding the
original lattice. The cut on the right can then be represented as a perfect matching
on the auxiliary graph as is shown in the middle panel of
Fig.~\ref{fig:kasteleyn}. Here, vertices that do not have cut lines adjacent to them
will have all four vertices of the associated Kasteleyn city matched by the internal
edges, such that after contracting back the Kasteleyn cities to regular vertices one
ends up with the graph shown on the right, that represents the cut in spin
language. To ensure that a MWPM corresponds to a minimum cut, we assign edge weights
in the auxiliary graph that are equal to the coupling $J_{ij}$ of the bond in the
original graph that is crossed by the bond in the auxiliary graph. For bonds in the
auxiliary graph that do not correspond to edges in the original graph, in particular
the internal bonds of Kasteleyn cities as well as bonds between the additional
external plaquettes, the weight is set to zero. Finally, a spin configuration
consistent with the loops on the dual graph found in this way is constructed by
flipping the spin orientation each time a loop line is crossed
\cite{thomas:07,gregor:07}.

\begin{figure*}
  \includegraphics[width=0.95\textwidth]{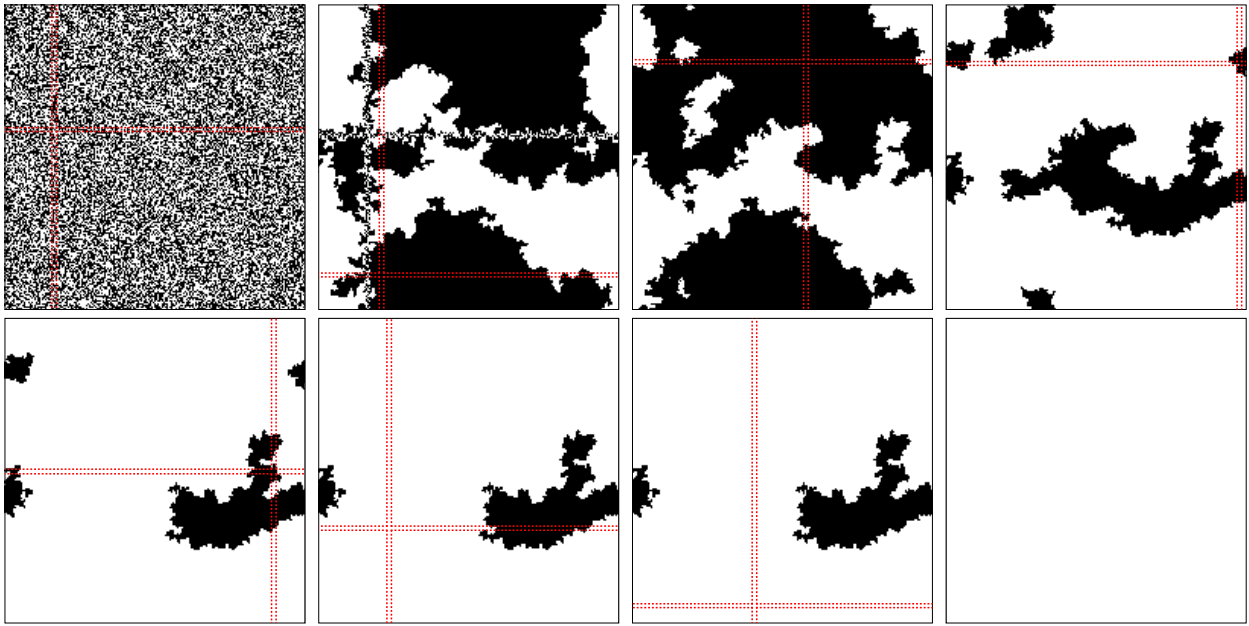}
  \caption{%
    Application of the windowing method to find a ground state of a sample with
    toroidal boundaries. Spins on white lattice sites are consistent with the ground-state
    orientation $s_i^0$, i.e., $s_i s_i^0 = +1$, black spins are oppositely oriented,
    i.e., $s_i s_i^0 = -1$. In a random initial
    configuration the spins have $s_i s_i^0 = \pm 1$ uniformly at random (top
    left). Exact ground states are found in windows of size $(L-2)\times (L-2)$
    placed at a random location (red dotted lines), with the remaining spins acting as fixed
    boundaries. After a few iterations all spins have the ground-state orientation
    (bottom right).
    \label{fig:window-example}
  }
\end{figure*}

As the auxiliary graph used here has only $4(L+1)^2$ vertices and
$6(L+1)^2+2L(L-1) = 8L^2+10L+6$ edges (for periodic-free boundaries) as compared to
the $O(L^2)$ vertices and $O(L^4)$ edges of Bieche's approach \cite{bieche:80a}, it
is significantly more efficient and, due to the smaller storage requirements, this
approach allows to treat much larger systems sizes. In practice, we use the Blossom V
implementation introduced in Ref.~\onlinecite{kolmogorov:09} to perform the MWPM
calculations. The method can be easily generalized to other planar graphs, for
instance $L\times L$ graphs with periodic boundaries in one direction. In this case,
the two additional lines of external plaquettes in either the horizontal or vertical
direction can be removed, otherwise the algorithm proceeds in the same way. A
generalization to non-planar graphs is not possible, however, as then the one-to-one
mapping between solutions of the MWPM problem and ground states of the spin system
breaks down \cite{thomas:07}: if the solution to the MWPM leads to loops that wrap
around the lattice it is possible to find an odd number of loop lines in a given row
or column of the lattice. In this case, it is not possible to find a spin
configuration that is consistent with the lines.

\subsection{Windowing technique for toroidal systems}
\label{sec:window}

Such a configuration with an odd number of line segments in a given row or column of
an $L\times L$ system with fully periodic boundaries can be repaired by changing the
boundary conditions in the corresponding direction from periodic to antiperiodic,
corresponding to an extra loop wrapping around the lattice, thus resulting in an even
number of lines again. In this sense, as discussed in Ref.~\onlinecite{thomas:07},
the approach outlined above finds an {\em extended ground state\/} for a system where
the boundary conditions are added to the dynamical degrees of freedom. While this can
be quite useful, it is not immediately applicable to the calculation of defect
energies and domain walls, where specific, fixed boundary conditions need to be
applied.

Nevertheless, a method for finding ground states for a fixed choice of periodic
boundary conditions can be constructed from the MWPM approach outlined above, as we
will now show. To achieve this, we successively determine exact ground states in
square windows of size $L'\times L'$, $L' \le L$, with free boundary conditions,
while the spin configuration outside of the window remains unchanged. By moving this
window randomly over the full $L\times L$ lattice, the exact ground state is
typically found after a moderate number of iterations. The sequence is started by
initializing the system in a random spin configuration $\{s_i\}$. The origin of the
window is then chosen randomly at one of the lattice sites, and the exact ground
state of the spins inside of the window is determined using MWPM, subject to the
additional constraint of a layer of fixed spins surrounding it. These spins are fixed
by placing very strong bonds with couplings $J_\mathrm{strong}$ between them that
cannot be broken in the solution of the MWPM, for instance by choosing
$|J_\mathrm{strong}| > \sum_{\langle ij\rangle}|J_{ij}|$. We choose
$J_{ij} = +|J_\mathrm{strong}|$ for parallel spins along the boundary of the window
and $J_{ij} = -|J_\mathrm{strong}|$ for antiparallel ones to ensure that these spins
do not change their relative orientation as a result of the MWPM run. This setup is
illustrated in Fig.~\ref{fig:windowing_technique}.

\begin{table}[tb!]
  \caption{%
    The average probability $\overline{P}_{n}$ of finding the ground
    state (success probability) for $20\leq L \leq 1000$,
    and for different numbers $n$ of iterations. Results are averaged over
    100 disorder realizations.
  \label{tab:parameters}
}
\begin{ruledtabular}
  \begin{tabular}{lcccccc}
    $L \backslash n  $ &   5     &    10     &   15   &   20   &  25  & 30 \\
    \hline
    20    &  0.276  &   0.561   &  0.671 &   0.728  &  0.762  &  0.782  \\
    50    &  0.317  &   0.603   &  0.705 &   0.756  &  0.790  &  0.805  \\
    80    &  0.315  &   0.592   &  0.700 &   0.752  &  0.783  &  0.806  \\
    100   &  0.315  &   0.594   &  0.700 &   0.745  &  0.779  &  0.789  \\
    150   &  0.326  &   0.611   &  0.714 &   0.768  &  0.797  &  0.821  \\
    200   &  0.323  &   0.610   &  0.712 &   0.765  &  0.792  &  0.814  \\
    350   &  0.340  &   0.628   &  0.729 &   0.789  &  0.822  &  0.833  \\
    500   &  0.317  &   0.589   &  0.683 &   0.740  &  0.771  &  0.801  \\
    700   &  0.329  &   0.612   &  0.723 &   0.770  &  0.782  &  0.818  \\
    1000  &  0.322  &   0.609   &  0.713 &   0.764  &  0.779  &  0.807  \\
  \end{tabular}
\end{ruledtabular}
\end{table}

As the spins at window boundaries are fixed and the resulting constraint optimization
problem is solved exactly, each iteration of the windowing method decreases the
energy of the total system or leaves it invariant. We observe convergence of the
method after a moderate number $n$ of iterations. The process is illustrated in
Fig.~\ref{fig:window-example}, where we display the overlap $s_i s_i^0$ with the
exact ground state $s_i^0$ for an example disorder configuration of linear size
$L=200$ with Gaussian couplings starting from a random initial spin configuration. It
is seen how even the first optimization with a window of size $L'=L-2=198$ leaves
only a single (large) cluster excitation over the ground state. As is seen from the
following panels, such excitations can only be fully relaxed if the window does not
intersect them. Hence the time until convergence is a random variable. To determine a
good set of parameters we performed test runs for different sizes $L$ and $L'$ of the
system and the window, respectively, and with a varying number of iterations. The
results show that the necessary number of iterations depends both on $L'$ and the
initial spin configuration, such that larger $L'$ needs smaller $n$, and if the
initial spin configuration is changed, $n$ will also change. As is intuitively
plausible, we find best results for the largest windows, and so we fixed the window
size to its maximum $L'=L-2$ for all runs.  To decide whether a given run arrives in
one of the ground states, we compared against exact results for system sizes
$L\le 100$ produced by the branch-and-cut method implemented in the spin-glass server
\cite{sgs}. For larger system sizes we used the lowest energy found in a sequence of
independent runs as an estimate of the ground-state energy and measured the success
probability $P_{n}(\{J_{ij}\})$ as the proportion of runs that ended in this
lowest-energy state found or in the exact ground-state for the system sizes treated
by the spin-glass server. The resulting success probability data, estimated from
between $250$ ($L\ge 700$) to $2000$ ($L\le 150$) runs for different initial spin
configurations for each disorder realization, is collected in Table
\ref{tab:parameters}. As is clearly seen, the success probabilities are rather high
such that for $n = 20$, for instance, they are consistently above $70\%$. There is
almost no size dependence of the average success probability $\overline{P}_{n}$, so
the hardness of finding ground states for the fully periodic torus lattices with the
proposed method does not increase with system size.

Still, from the data presented in Table \ref{tab:parameters}, it is clear that not
every run of the windowing method converges to the ground state. To further increase
the success probability of the method, we use repeated runs and pick the lowest
energy found there \cite{weigel:06b}. If the success probability for a given sample
in runs of ${n}$ iterations is $P_{n}(\{J_{ij}\})$, then the probability of finding
the ground state at least once in $m$ independent runs is
\begin{equation}
  P_s(\{J_{ij}\})=1-[1-P_{n}(\{J_{ij}\})]^m,
  \label{eq:success}
\end{equation}
and this can be tuned arbitrarily close to unity by increasing $m$. If we set a
desired success probability of, say, $P_s = 0.999$, we can use Eq.~\eqref{eq:success}
to determine the required number $m$ of repetitions. For each realization we hence
find
\[
m(\{J_{ij}\}) = \log[1-P_s]/\log[1-P_{n}(\{J_{ij}\})].
\]
In Table \ref{tab:parameters2} we show the values of $\overline{m}$ averaged over 100
disorder realizations as a function of $L$ and $n$. Clearly, the dependence on system
size is weak. The total computational effort of such repeated runs is proportional to
$m \times n$. From the values of $n$ tested in Table \ref{tab:parameters2}, this
effort is found to be minimal for $n = 10$, and we use $m=8$ repetitions independent
of system size to find the exact ground state in approximately $99.9\%$ of the
samples. As an additional protection against potential outliers we demand that the
lowest-energy state found in these $m=8$ runs must have occurred at least three out
of these $8$ times. If this is not the case, another $8$ runs are performed etc. This
adds only a tiny fraction of extra average runtime, but it will be able to catch a
few of the $0.01\%$ of samples where the ground state would otherwise not be
found. As a test, we applied this combined technique to the samples for $L\le 100$
where the exact ground-state energy is known and it arrived in a ground state in all
cases.

\begin{table}[tb!]
  \caption{%
    The average number $\overline{m}$ of repetitions required according to
    Eq.~\eqref{eq:success} for runs of the windowing technique with $n$ random
    placements of the window per run to ensure an overall success probably of $P_s =
    0.999$.
    \label{tab:parameters2}
  }
  \begin{ruledtabular}
    \begin{tabular}{lcccccc}
      $L \backslash n  $ &   5     &    10     &   15   &   20   &  25  & 30 \\
      \hline
      20    &  23.5  &   9.3   &  6.9 &   5.8  &  5.3  &  4.9  \\
      50    &  19.7  &   8.2   &  6.2 &   5.3  &  4.7  &  4.5  \\
      80    &  20.2  &   8.6   &  6.4 &   5.4  &  4.9  &  4.4  \\
      100   &  20.0  &   8.5   &  6.4 &   5.6  &  4.9  &  4.5  \\
      150   &  19.2  &   8.1   &  6.0 &   5.0  &  4.5  &  4.0  \\
      200   &  19.2  &   8.3   &  6.1 &   5.3  &  4.7  &  3.6  \\
      350   &  18.4  &   7.4   &  5.8 &   4.8  &  3.9  &  3.8  \\
      500   &  21.2  &   8.5   &  7.0 &   6.0  &  5.0  &  4.6  \\
      700   &  20.3  &   7.9   &  6.3 &   5.8  &  4.6  &  5.0  \\
      1000  &  20.0  &   8.5   &  6.4 &   5.1  &  5.0  &  4.5  \\
    \end{tabular}
  \end{ruledtabular}
\end{table}
 
\subsection{Performance of the algorithm}
\label{sec:performance}

It is interesting to see how the matching based on Kasteleyn cities for planar
instances as well as the windowing method outlined above for toroidal graphs fare in
computational efficiency as compared to the more general approaches implemented in
the spin-glass server \cite{sgs}. The run times in seconds on standard hardware are
shown for periodic-free boundary conditions (PFBC) and for periodic-periodic
(toroidal) boundaries (PPBC) as compared to the corresponding results of the
spin-glass server for system sizes $L\le 100$ in Table \ref{windowing_time}. For PFBC
the matching approach is always much faster than the method used by the spin-glass
server, which is based on a modified exact numeration technique known as
branch-and-cut. For PPBC the windowing technique introduces a certain overhead,
such that a crossover is observed with branch-and-cut being faster for
$L \lesssim 20$ and the windowing method winning out for $L\gtrsim 20$.

\begin{table}[tb!]
  \caption{%
    Average run time (in seconds) for determining a ground state of samples with
    periodic-free boundaries (PFBC) and periodic-periodic boundaries (PPBC),
    respectively, using the minimum-weight perfect matching (MWPM) approach based on
    Kasteleyn cities for PFBC and the windowing technique (WT) for PPBC as compared to
    the times reported by the spin-glass server (SGS) on the same samples.
  }
  \begin{ruledtabular}
    \begin{tabular}{lllll}
      % \hline
      $L$ & \multicolumn{2}{c}{PFBC}  & \multicolumn{2}{c}{PPBC}  \\
      \cline{2-3}       \cline{4-5} 
          & SGS & MWPM &  SGS & WT \\
      \hline
      8   &  0.00228 &  0.000203 &  0.00560 &  0.02468  \\
      10  &  0.01330	&  0.000424 &  0.01950 &  0.04462  \\
      20  &  0.18330	&  0.002361 &  0.22820 &  0.19119  \\
      50  &  3.38740	&  0.024184 &  3.93040 &  2.18788  \\
      80  &  31.0738	&  0.069104 &  35.7004 &  6.42005  \\
      100 &  150.218	&  0.115761 &  189.501 &  9.81247  \\
      % \hline
    \end{tabular}
  \end{ruledtabular}
  \label{windowing_time}
\end{table}

\begin{figure}
  \begin{center}
    \includegraphics[width=0.95\columnwidth]{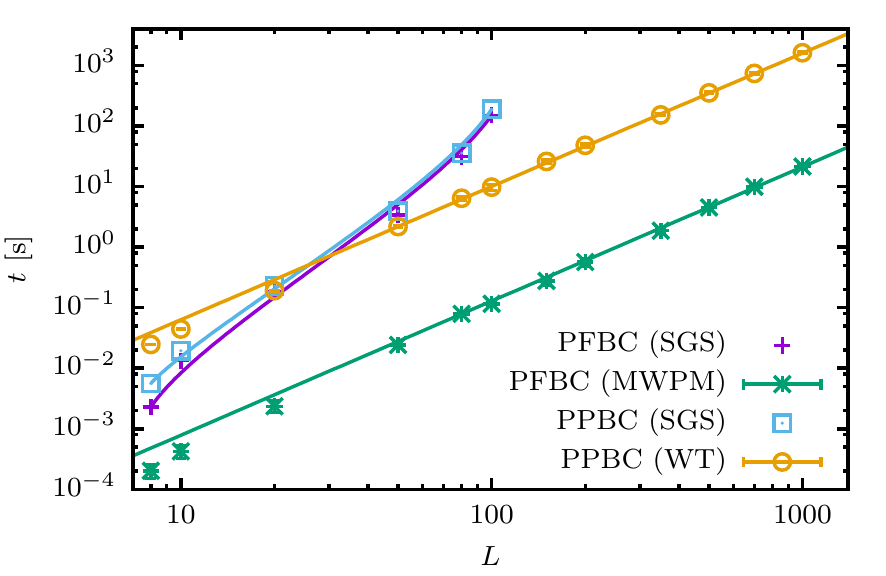}
    \caption{(Color online) Average time $t$ per sample to determine ground states of
      systems with PFBC and PPBC for $L\times L$ samples using the minimum-weight
      perfect matching (MWPM) method for periodic-free samples (PFBC), the windowing
      technique (WT) for periodic-periodic samples (PPBC), and the spin-glass server
      (SGS), respectively. The straight lines are fits of the form
      \eqref{eq:power-law} to the data, whereas the lines for the SGS data are just
      interpolations to guide the eye.}
    \label{fig:run_time}
  \end{center}
\end {figure}

The scaling of run times with system size is illustrated in
Fig.~\ref{fig:run_time}. The algorithm of the spin-glass server utilized here is
based on branch-and-cut \cite{liers:04}, which corresponds to a combination of a
cutting plane technique with the iterative removal of branches of the search tree
that cannot contain a solution. While this approach is quite efficient, and
outperforms other exact methods for hard problems, its run-time still scales
exponentially with system size. The super-polynomial behavior is clearly seen in the
doubly logarithmic representation of Fig.~\ref{fig:run_time}. For the matching
approach for PFBC, the implementation used here has $O(L^6)$ worst-case scaling
\cite{kolmogorov:09}. As the straight line indicates, we indeed see clear power-law
behavior, but the average run times probed here increase much more gently with system
size. A power-law fit of the form
\begin{equation}
  \label{eq:power-law}
  t(L) = A_t L^{\kappa}
\end{equation}
to the data yields $\kappa = 2.22(2)$, so the scaling is only slightly worse than
linear in the volume in the considered range of system sizes.

Finally, for the windowing technique built on top of MWPM for the PPBC samples, we
find an overhead that is to a very good approximation independent of system size,
such that calculations for PPBC are by a factor of $80$ more expensive that those for
samples with PFBC for the chosen confidence level of $P_s = 0.999$, corresponding to
the $n = 10$ iterations and $m=8$ repetitions. A fit of the form
\eqref{eq:power-law} to the data for PPBC yields $\kappa = 2.20(2)$, perfectly
consistent with the results for PFBC. The ratio of amplitudes $A_t$ is estimated as
$A_t = 83\pm 12$, consistent with the expected value of slightly above $80$ resulting
from the additional requirement of a threefold occurrence of the ground state.

\section{Results for Gaussian couplings\label{sec:gauss}}

For the Gaussian distribution \eqref{eq:gaussian} the set of couplings for which
exact degeneracies occur is expected to be of zero measure. The present techniques
based on matching hence directly yield the correct distribution of states at zero
temperature.

\subsection{Ground-state energies}

The average ground-state energy per spin, $\langle e(L)\rangle_J$, depends on the
coupling distribution. Additionally, we expect finite-size corrections which in turn
are sensitive to the boundary conditions employed
\cite{bouchaud:03,hartmann:04a,weigel:06c}. Following Ref.~\onlinecite{hartmann:04a}
one expects a Wegner correction exponent $\omega(d) = (6-d)+\cdots$ to leading order,
whereas numerically one finds \cite{hasenbusch:08} $\omega \approx 1.0$ for Ising
spin glasses in $d=3$ and $\omega\approx 0.75$ for $d=2$. \cite{fernandez:16} As then
$-(d-\theta)+\omega \approx -3.03$ in two dimensions, this implies that non-analytic
corrections are substantially suppressed against the leading analytic ones in this
quantity. We hence assume the following general form for the size dependence of the
average ground-state energy,
\begin{equation}
  \begin{split}
    \langle e(L)\rangle_J = & e_\infty + A_E L^{-(d-\theta)} + C_E L^{-1} + \\
    & D_E L^{-2} + E_E L^{-3} + \ldots.
  \end{split}
  \label{eq:energy_fit_form}
\end{equation}
The presence of a term proportional to $L^{-(d-\theta)}$ follows from standard
arguments about the scaling of the correlation length and the free-energy density
\cite{privman:privman}, taking additionally into account that for a $T=0$ critical
point the $1/\beta^2$ prefactor in the relation $e = (-1/\beta^2)\d(\beta f)/\d T$ is
critical, as well as making use of the relation $\nu = -1/\theta$.
\cite{hartmann:04a} Although this derivation should apply for any $T=0$ critical
point, for the spin glass it is tempting to attribute the occurrence of the
$L^{d-\theta}$ term to the presence of domain-wall defects that are trapped in the
system due to periodic boundary conditions. In Ref.~\onlinecite{hartmann:04a} it is
suggested to reduce the number of parameters in Eq.~\eqref{eq:energy_fit_form} by
considering the energy $\hat{e}(L)$ per bond instead of the energy $e(L)$ per
site. If one assumes that depending on the boundary conditions this quantity has a
$1/L$ correction for any free edge and a $1/L^2$ correction for any corner, for the
square lattice with its two bonds per site we expect
\[
  2\langle \hat{e}(L)\rangle_J= e_\infty + \hat{A}_E L^{-(d-\theta)} + \hat{C}_EL^{-1} +
  \hat{D}_EL^{-2}
\]
up to higher-order corrections. For free-free boundaries, one has $E(L) = L^2 e(L) =
(2L^2-2L)\hat{e}(L)$ and hence
\begin{equation}
  \label{eq:energy_ffbc}
  \begin{split}
    \langle e(L)\rangle_J = & e_\infty + \hat{A}_E L^{-(d-\theta)} + (\hat{C}_E-e_\infty) L^{-1}\\
    & +(\hat{D}_E-\hat{C}_E) L^{-2}-\hat{D}_E L^{-3},
  \end{split}
\end{equation}
where a term of order $L^{-(d-\theta)-1}$ which for $\theta < 0$ is asymptotically
smaller than $1/L^3$ has been neglected. This is of the form of
Eq.~\eqref{eq:energy_fit_form}, but with the $1/L^3$ term merely being produced by
the $1/L^2$ correction in $\hat{e}(L)$, such that there are only five fit parameters
in \eqref{eq:energy_ffbc} as compared to six parameters in
Eq.~\eqref{eq:energy_fit_form}.  For periodic-free boundaries there is a free edge
but no corners, such that $\hat{D}_E=0$ and $E(L) = (2L^2-L)\hat{e}(L)=L^2e(L)$, and
we find
\begin{equation}
  \label{eq:energy_pfbc}
  \begin{split}
    \langle e(L)\rangle_J = & e_\infty + \hat{A}_E L^{-(d-\theta)} + (\hat{C}_E-e_\infty/2) L^{-1}\\
    & -(\hat{C}_E/2) L^{-2},
  \end{split}
\end{equation}
where again a term proportional to $L^{-(d-\theta)-1}$ was omitted. For
periodic-periodic boundaries, on the other hand, one should have
$\hat{C}_E = 0 = \hat{D}_E$, and hence only a correction proportional to
$L^{-(d-\theta)}$. We will test the validity of these assumptions for our data below.

Beyond the mean ground-state energy, it is interesting to study the shape of the
energy distribution over different disorder samples. It has been shown in
Ref.~\onlinecite{bouchaud:03}, based on results of Wehr and Aizenman \cite{wehr:90},
that the width of this distribution scales as $L^{\Theta_f}$ with $\Theta_f = -d/2$.
Below, we investigate the distribution shape by direct inspection and by analyzing
the scaling of its kurtosis defined by
\begin{equation}
  \label{eq:kurtosis}
  \operatorname{Kurt}[e] = \frac{\langle (e-\langle e\rangle_J)^4\rangle_J}
  {[\langle (e-\langle e\rangle_J)^2\rangle_J]^2}
\end{equation}
with system size, where $\operatorname{Kurt}[\cdot] = 3$ for a Gaussian distribution.

\subsection{Domain-wall calculations}

\begin{table}[tb!]
  \caption{The number of disorder realizations for different boundary conditions,
    coupling distributions and system sizes.}
  \begin{ruledtabular}
    \begin{tabular}{l c c c}
      $L$   &  PFBC Gaussian   & PPBC Gaussian    &    PFBC bimodal      \\
      \hline                                                           
      8     &  $1\times10^6$   & $1\times10^5$    &  $1\times10^5$       \\
      10    &  $1\times10^6$   & $1\times10^5$    &  $1\times10^5$       \\
      20    &  $1\times10^6$   & $1\times10^5$    &  $1\times10^5$       \\
      30    &  $1\times10^6$   & $1\times10^5$    &  $1\times10^5$       \\
      40    &  $1\times10^6$   & $1\times10^5$    &  $1\times10^5$       \\
      50    &  $1\times10^6$   & $1\times10^5$    &  $1\times10^5$       \\
      80    &  $1\times10^6$   & $8\times10^4$    &  $1\times10^5$       \\
      100   &  $1\times10^6$   & $8\times10^4$    &  $1\times10^5$       \\
      150   &  $1\times10^6$   & $1\times10^5$    &  $1\times10^5$       \\
      200   &  $1\times10^6$   & $5\times10^4$    &  $8\times10^4$       \\
      350   &  $5\times10^5$   & $5\times10^4$    &  $8\times10^4$       \\ 
      500   &  $5\times10^5$   & $3\times10^4$    &  $5\times10^4$       \\ 
      700   &  $5\times10^5$   & $1\times10^4$    &  $3\times10^4$       \\ 
      1000  &  $3\times10^5$   & $1\times10^4$    &  $1\times10^4$       \\
      1500  &  $1\times10^5$   & $7\times10^3$    &  $5\times10^3$       \\
      2000  &  $5\times10^4$   & $1\times10^3$    &  $3\times10^3$       \\
      3000  &  $3\times10^4$   & $640        $    &  $1505       $       \\
      4000  &  $2\times10^4$   &                  &                      \\
      5000  &  $3\times10^3$   &                  &                      \\
      7000  &  $400        $   &                  &                      \\   
      8000  &  $455        $   &                  &                      \\   
      10000 &  $265        $   &                  &                      \\   
    \end{tabular}
  \end{ruledtabular}
  \label{tab:samples}
\end{table}

The analysis of defect energies provides a convenient way of studying the stability
of the ordered phase. In the most common approach one inserts system-spanning domain
walls into the system by a suitable change of boundary conditions
\cite{banavar:82a}. The energy of such excitations scales as a power of their linear
size \cite{bray:84},
\begin{equation}
    E_{\mathrm{def}} \propto L^{\theta}, \label{eq:defE}
\end{equation}
where the spin-stiffness exponent $\theta$ depends on the symmetries of the model as
well as the lattice dimension $d$. In a simple generalization of Peierls' argument
for the stability of the ferromagnetic phase, one concludes that a spin-glass phase
is stable against thermal fluctuations up to some $T_c>0$ if $\theta > 0$ and
unstable for $\theta < 0$, with $\theta = 0$ denoting the marginal case. The
conceptually most direct way of inserting a domain-wall excitation is to compute a
ground-state for free boundaries in, say, the $x$ direction as a reference and to
then fix the boundary spins along the $x$ boundary in opposite relative orientations
as compared to this state for a second ground-state calculation. The excess energy in
the second run corresponds to the energy contained in the domain wall. This setup is
sometimes referred to as domain-wall boundary condition
\cite{hartmann:01a,carter:02a}. An alternative proposed initially by Banavar
\cite{banavar:82a} uses the difference between the ground-state energies for periodic
and for antiperiodic boundaries in $x$ direction. The resulting value of
$\Delta E = E_\mathrm{P}-E_\mathrm{AP}$ is potentially the difference of energies of
two configurations with such domain walls as the periodicity of both P and AP
boundaries can force a domain wall into the system \cite{kosterlitz:99a,weigel:05f},
but this difference is found to nevertheless scale with the same stiffness exponent
as for domain-wall boundaries \cite{carter:02a}.

\begin{figure}
  \begin{center}
    \includegraphics[width=0.95\columnwidth]{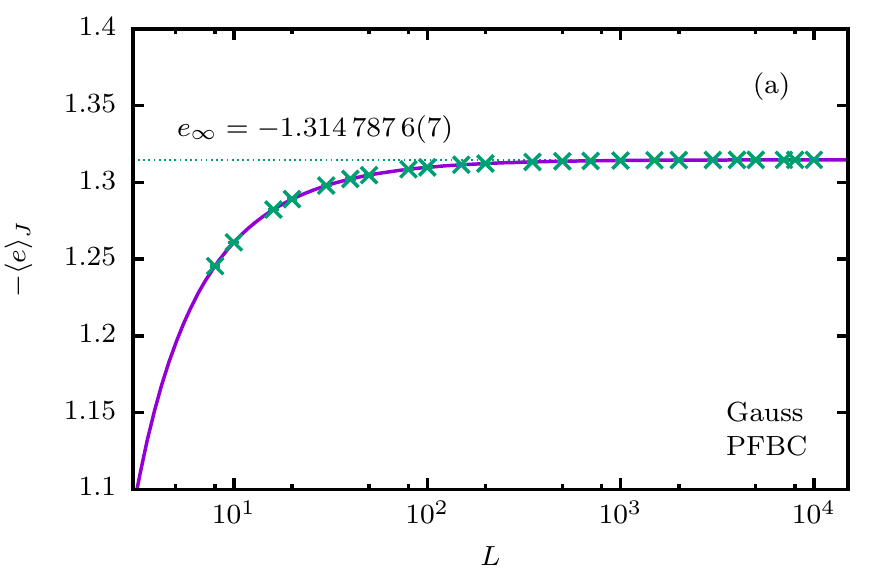}
    \includegraphics[width=0.95\columnwidth]{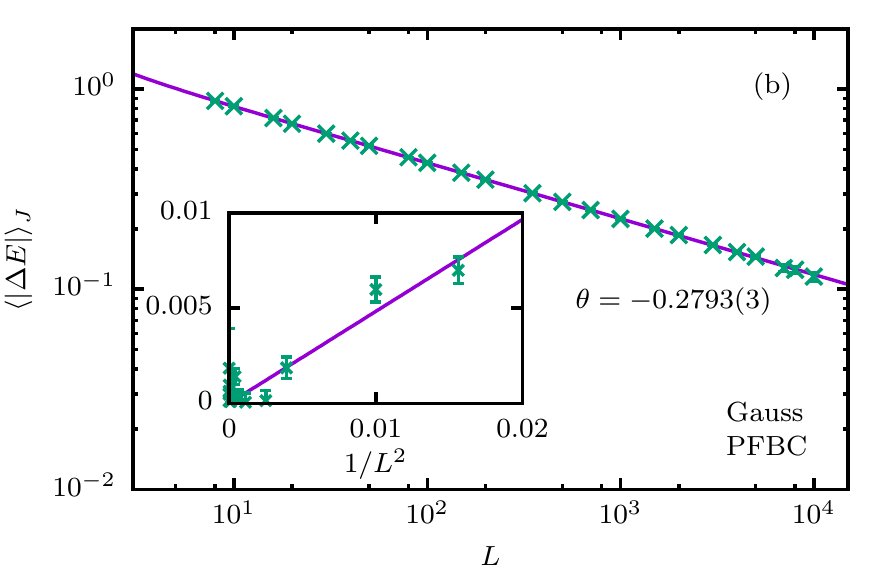}
    \includegraphics[width=0.95\columnwidth]{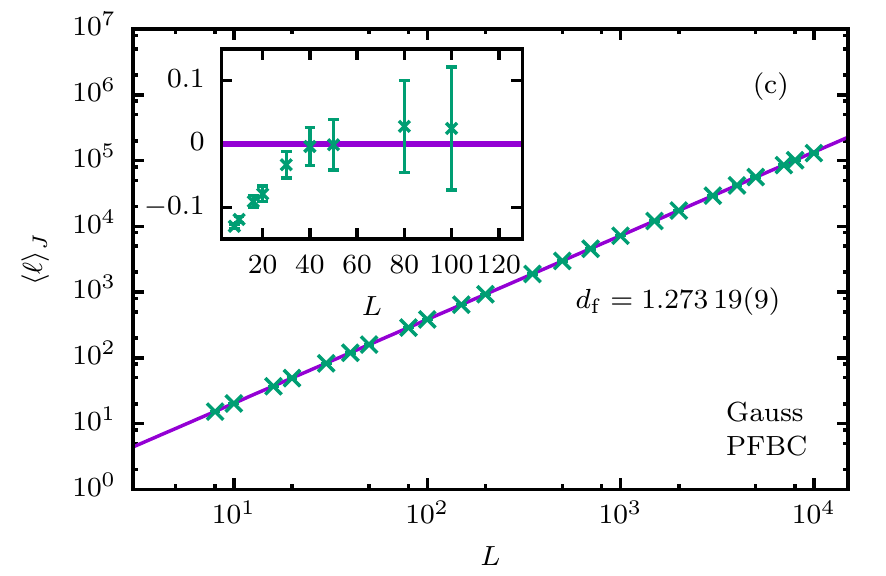}
    \caption{%
      (Color online) (a) Disorder-averaged ground-state energy per site
      $\langle e\rangle_J = \langle\bar{E}/L^2\rangle_J$ for PFBC and Gaussian
      couplings together with a fit of the form \eqref{eq:energy_pfbc} to the data in
      the range $L=10,\ldots,10\,000$.  (b) Average defect energies
      $\langle|\Delta E|\rangle_J$ for the same system as calculated from the
      difference in ground-state energies between periodic and antiperiodic boundary
      conditions in the $x$ direction.  The points show our data for
      $8 \leq L \leq 10\,000$ and the solid line represents a fit of the form
      $\langle|\Delta E|\rangle_J(L) = A_\theta L^\theta + C_\theta/L^2$ to the data.
      The inset shows the correction
      $\langle|\Delta E|\rangle_J(L) - A_\theta L^\theta$ plotted against $1/L^2$
      illustrating that this single term describes the corrections very well. (c)
      Average length $\ell$ of the domain-wall in the overlap of ground states for
      periodic and antiperiodic boundaries in $x$ direction and free boundaries in
      $y$ direction (PFBC boundaries). The line shows a fit of the functional form
      $\langle \ell\rangle_J = A_\ell L^{d_\mathrm{f}}$ to the data for
      $L\ge L_\mathrm{min} = 40$. The inset shows a blow-up of the deviations for
      small $L$.}
    \label{fig:pfbc_G}
  \end{center}
\end {figure}

For calculations based on MWPM alone one needs to apply free boundaries in $y$
direction in order to ensure planarity of the lattice. With the help of the windowing
technique it is also possible to implement this procedure for samples with
periodic-periodic boundaries, however. In general we expect the leading scaling to be
accompanied by scaling corrections of the form \cite{weigel:06c}
\begin{equation}
  \label{eq:defect-energy-corrected}
  \langle|\Delta E(L)|\rangle_J(L) = A_\theta L^{\theta} (1+B_\theta L^{-\omega})
  +\frac{C_\theta}{L}+\frac{D_\theta}{L^2}+\cdots,
\end{equation}
where $\omega$ denotes the leading corrections-to-scaling exponent, and $1/L$ and
$1/L^2$ are analytic corrections \cite{privman:privman}. For the setup with
domain-wall boundary conditions significantly stronger corrections have been observed
than for the P-AP situation \cite{carter:02a} and we hence concentrate on the latter
approach here.

Apart from the energy density of domain walls or droplet boundaries another
contentious question is that of the geometric nature of excitations in spin
glasses. While it is not ultimately clear whether droplets or domain walls are the
fundamental objects in this system or rather some more esoteric form of excitations
such as sponges exist \cite{palassini:00,krzakala:00,newman:01}, it is interesting to
see whether domain-walls are stochastically fractal objects and if the corresponding
fractal dimension $d_\mathrm{f} < d$ or rather domain walls can be space-filling
\cite{marinari:00}. We determined the domain wall as the set ${\cal D}$ of all dual
bonds for which
\begin{equation}
  [J_{ij} s_i s_j]^{(\mathrm{P})} [J_{ij} s_i s_j]^{(\mathrm{AP})} < 0.
  \label{eq:dw-def}
\end{equation}
The inclusion of the couplings $J_{ij}$ in the product takes care of the fact that
across the edge where the boundary condition is changed from P to AP the spins will
be in different relative orientation before and after the change, but this is merely
a consequence of the flip $J_{ij} \to -J_{ij}$ of the couplings there and should not
be counted as a part of the induced domain wall. We denote by $\ell$ the number of
(dual) edges in the set ${\cal D}$. Following the usual box-counting argument,
scaling according to $\langle \ell\rangle_J \sim L^{d_\mathrm{f}}$ defines the
domain-wall fractal dimension $d_\mathrm{f}$. As for the defect energies we
anticipate the presence of corrections, leading to the scaling form
\begin{equation}
  \langle\ell\rangle_J(L) = A_\ell L^{d_\mathrm{f}} (1+B_\ell L^{-\omega})
  +\frac{C_\ell}{L}+\frac{D_\ell}{L^2}+\cdots.
\end{equation}

\subsection{Periodic-free boundaries}

For the periodic-free setup (PFBC) we used the MWPM approach for periodic and
antiperiodic boundaries in $x$ direction and system sizes ranging from $L=8$ up to
$L=10\,000$. For $L\le 350$ we generated $10^6$ disorder configurations, while for
larger systems the number of replicas is gradually reduced down to about $300$ for
$L=10\,000$, see the details collected in Table \ref{tab:samples}. We used the MIXMAX
random number generator \cite{savvidy:91,savvidy:15} which has provably good
statistical properties and also passes all of the tests in the suite TestU01
\cite{lecuyer:07}. As an additional check in view of the high-precision nature of the
present study, part of our calculations were repeated with Mersenne twister
\cite{matsumoto:98}. All results were found to be perfectly consistent within error
bars.

\begin{figure}
  \begin{center}
    \includegraphics[width=0.95\columnwidth]{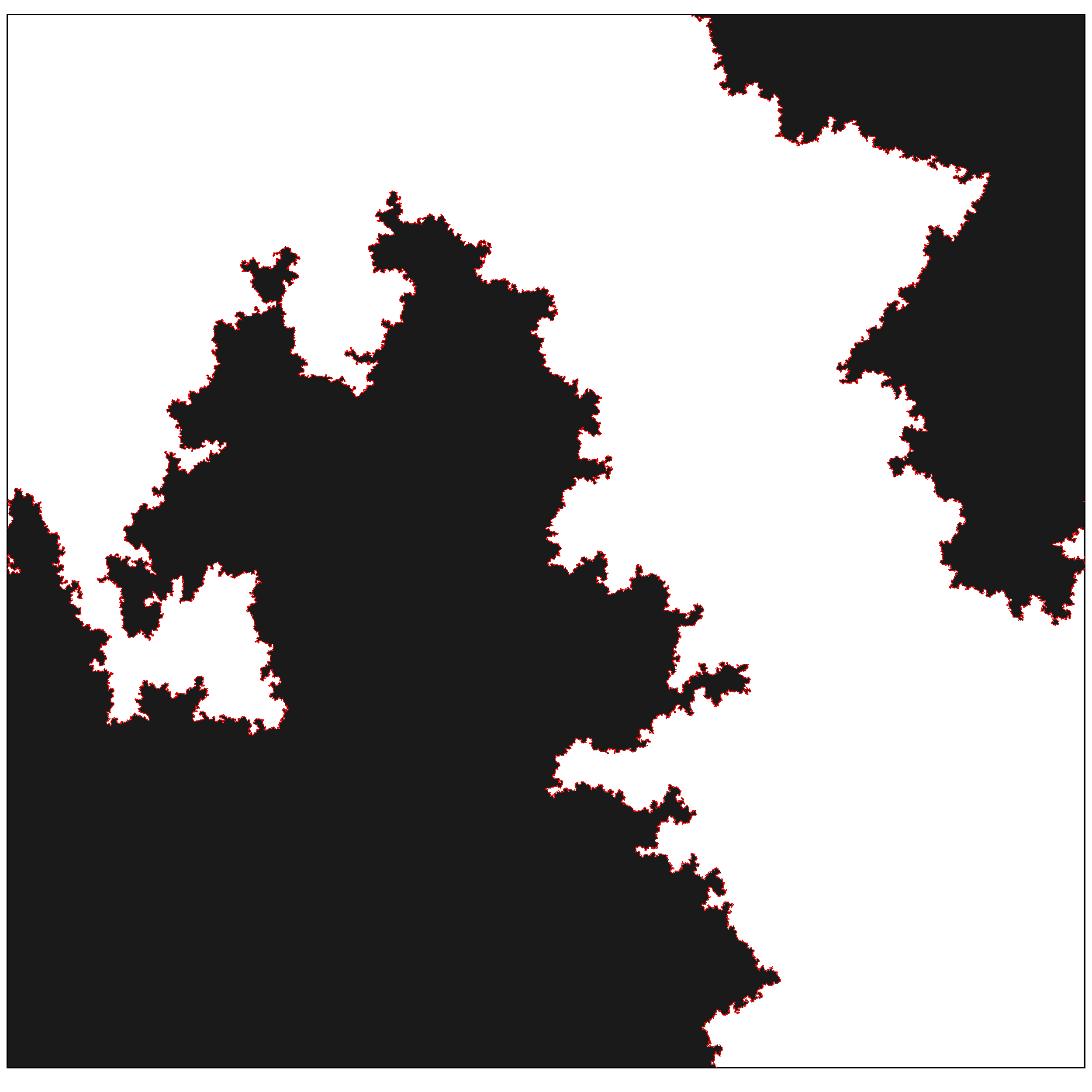}
    \caption{%
      (Color online) Overlap configuration of the ground states for P and AP
      boundaries for a $L=10\,000$ disorder realization of the PFBC Gaussian
      system. The red line demarcates the domain wall which traverses
      $\ell = 233\,141$ dual links.  }
    \label{fig:sampleDW}
  \end{center}
\end {figure}

We start by considering the ground-state energies. Here, we use the results for both
P and AP boundary conditions. They differ from each other, on average, by far less
than the statistical errors would suggest, but this is due to the fact that for each
sample both energies are highly correlated. For studying the average ground-state
energy, we hence calculated the average $\bar{E} =(E_\mathrm{P}+E_\mathrm{AP})/2$ and
estimated statistical errors for $\langle \bar{E}\rangle_J$ through the variation
over disorder samples. As the data in panel (a) of Fig.~\ref{fig:pfbc_G} show, the
finite-size corrections to scaling are relatively small, with the result for $L=10$
only being about 4\% above the asymptotic value. Due to the large range of system
sizes and high statistics in disorder samples we get a stable result for the full
non-linear five parameter fit of the form \eqref{eq:energy_pfbc} to the data with a
quality-of-fit\footnote{$Q$ is the probability that a $\chi^2$ as poor as the one
  observed could have occurred by chance, i.e., through random fluctuations, although
  the model is correct \cite{young:12}.} of $Q=0.81$. For the asymptotic ground-state
energy we find
\[
e_\infty = -1.314\,787\,6(7),
\]
while the spin-stiffness exponent $\theta = -0.273(65)$ from this fit\footnote{Note
  that hence the form \eqref{eq:energy_pfbc} is found to describe the data perfectly
  well, in contrast to the corresponding form used in Ref.~\onlinecite{hartmann:04a},
  cf.\ Eq.~(22) there, which is not consistent with the equation derived here.}. If
we fix $\theta$ at the value $\theta = -0.2793$ found below from the defect energy
calculations for the PFBC boundaries, the asymptotic ground-state estimate $e_\infty$
is unaltered from the above value up to the given number of digits. On gradually
increasing $L_\mathrm{min}$ we find statistically consistent fits that, however,
become less and less stable as the number of degrees of freedom is reduced. The
resulting estimate of $e_\infty$ is unaltered within statistical errors.

Our data for the defect energies are shown in Fig.~\ref{fig:pfbc_G}(b). We find
scaling corrections to be small and a pure power-law fit without corrections yields a
quality-of-fit $Q = 0.37$ for $L \ge L_\mathrm{min} = 50$.  The corresponding
estimate of the stiffness exponent is $\theta = -0.2798(4)$.  Corrections can hence
only be clearly resolved for $L\lesssim 50$. There, we find that the data are very
well described by a single correction term proportional to $1/L^2$, cf.\ the inset of
Fig.~\ref{fig:pfbc_G}(b), where we show the residual contribution
$\langle |\Delta E|\rangle_J - A_\theta L^\theta$ plotted against $1/L^2$. Our
$\theta$ estimate from this fit is
\[
\theta = -0.2793(3)
\]
with $Q=0.16$ when including all lattice sizes. Gradually increasing $L_\mathrm{min}$
does not reveal any discernible drift in the estimate for $\theta$. Since we have one
free boundary one might have expected the presence of a $1/L$ correction, which is
clearly present in the ground-state energy itself according to the fit following
Eq.~\eqref{eq:energy_pfbc}. In the energy difference $\Delta E$, however, this
contribution cancels out since the couplings along the free edge are absent in both
samples. If we nevertheless include such a term in the fit, its amplitude is found to
be consistent with zero. We are not able to clearly resolve a Wegner correction
$\propto L^{-\omega}$, which is not surprising since as discussed above we expect it
to be clearly weaker than $1/L^2$.

We finally turn to the domain-wall length. Fig.~\ref{fig:sampleDW} shows a sample
configuration with $L=10\,000$ illustrating the meandering nature of the domain
wall. For the average domain-wall length we find very clean scaling for PFBC as is
seen from our data depicted in Fig.~\ref{fig:pfbc_G}(c). A fit of the pure power-law
form $\langle \ell\rangle_J = A_\ell L^{d_\mathrm{f}}$ yields a fit quality of
$Q=0.56$ for $L_\mathrm{min} = 40$. The corresponding estimate of the fractal
dimension is
\[
d_\mathrm{f} = 1.273\,19(9).
\]
The deviations from a pure power law visible for system sizes $L < 20$ are rather
small and not well described by a single correction term. We hence prefer to take
them into account by simply omitting data from the small-$L$ side instead of
performing corrected fits. On systematically varying $L_\mathrm{min}$ in these fits,
we find a drift only for $L_\mathrm{min} \le 30$ and mutually consistent results for
larger $L_\mathrm{min}$.

\begin{figure}
  \begin{center}
    \includegraphics[width=0.95\columnwidth]{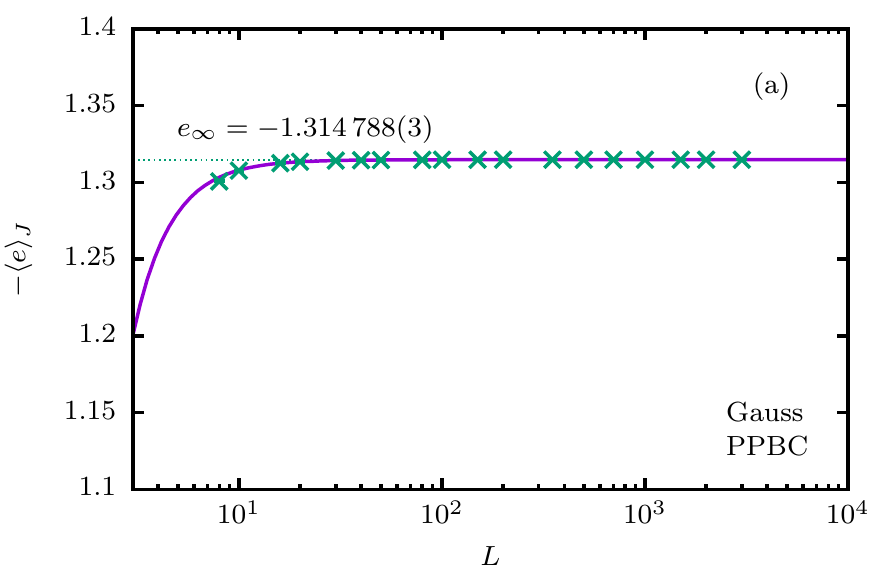}
    \includegraphics[width=0.95\columnwidth]{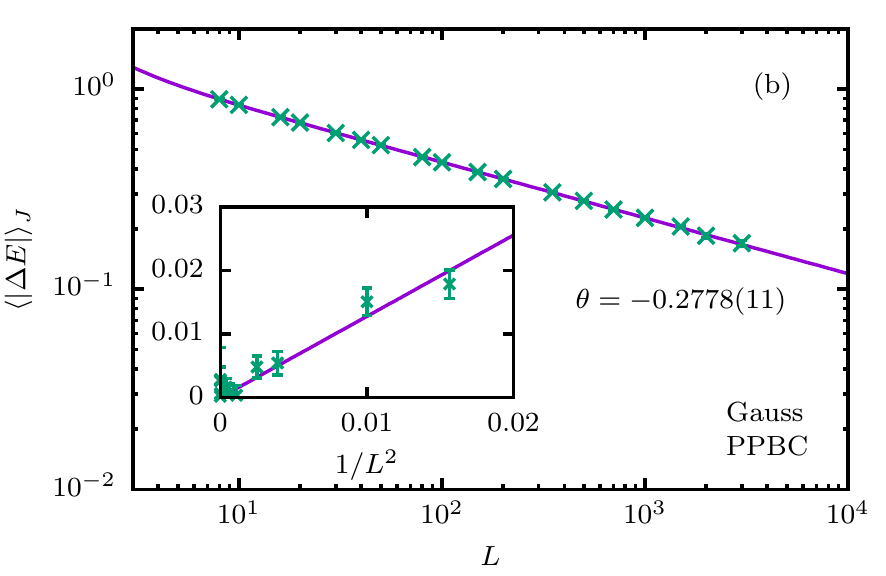}
    \includegraphics[width=0.95\columnwidth]{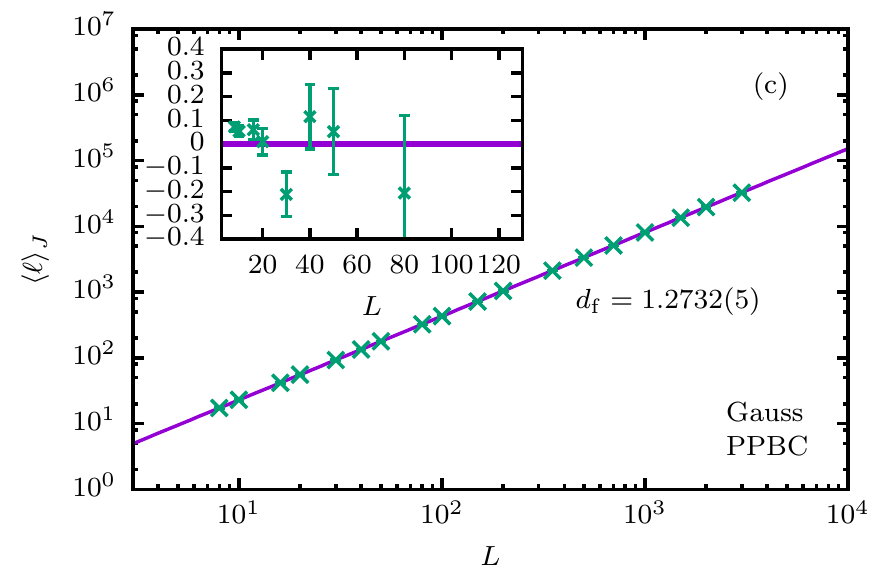}
    \caption{%
      (Color online) (a) Average ground-state energies for PPBC and Gaussian
      couplings together with a fit of the form \eqref{eq:energy_ppbc} to the data in the
      range $L \ge L_\mathrm{min} = 16$. (b) Scaling of defect energies for the Gaussian
      model with fully periodic boundary conditions. The solid line shows a fit of the form
      $\langle|\Delta E|\rangle_J(L) = A_\theta L^\theta + C_\theta/L^2$ to the data.  The
      inset shows the correction $\langle|\Delta E|\rangle_J(L) - A_\theta L^\theta$
      plotted against $1/L^2$ illustrating that this single term describes the corrections
      very well.  (c) Scaling of the length of the domain wall between P and AP ground
      states for the Gaussian PPBC case.  The solid lines shows a fit of the form $\langle
      \ell\rangle_J = A_\ell L^{d_\mathrm{f}}$ to the data with $L_\mathrm{min} = 40$. The
      inset shows a detail of the main plot for small $L$.
    }
    \label{fig:ppbc_G}
  \end{center}
\end {figure}

\subsection{Periodic-periodic boundaries}

For fully periodic or toroidal boundaries (PPBC) we use the windowing technique
discussed above in Sec.~\ref{sec:window} to find exact ground states in more than
99.9\% of the cases. Due to the increase in effort by the constant factor of $80$
resulting from the windowing technique, we reduced the maximum system size a bit and
considered lattices in the range $8\le L\le 3000$. Additionally, the number of
disorder realizations considered was reduced correspondingly, the exact numbers are
shown in Table~\ref{tab:samples}.

Our data for the ground-state energies for PPBC are shown in
Fig.~\ref{fig:ppbc_G}(a), illustrating that finite-size corrections in this case are
tiny, even much weaker than for the PFBC case. According to the discussion above, for
the ground-state energies we do not expect the presence of analytic corrections for
PPBC, and so we assume a scaling form
\begin{equation}
  \label{eq:energy_ppbc}
  \langle e\rangle_J = e_\infty + A_EL^{-(2-\theta)}.
\end{equation}
Fits of this form work very well and yield fit qualities of $Q>0.4$ for all
$L_\mathrm{min} \ge 10$. For $L_\mathrm{min} = 16$ we find
\[
  e_\infty = -1.314\,788(3)
\]
as well as $\theta = -0.35(14)$ and $A=1.51(65)$ with a good $Q=0.60$. This fit is
shown together with the data in panel (a) of Fig.~\ref{fig:ppbc_G}.

For the defect energies, the data again show clear power-law scaling with $L$, see
Fig.~\ref{fig:ppbc_G}(b). For $L \ge L_\mathrm{min} = 50$ we get an excellent fit
($Q=0.74$) for the pure power-law $\langle|\Delta E|\rangle_J = A_L L^\theta$ with
$\theta = -0.2778(14)$.  Regarding scaling corrections, it turns out that the size
range where they are visible is rather small. As the inset of
Fig.~\ref{fig:ppbc_G}(b) shows, corrections are well described by a single
$1/L^2$ term, consistent with the findings for the PFBC case. A corresponding fit for
$L_\mathrm{min} = 10$ yields high quality with $Q=0.92$ and
\[
  \theta = -0.2778(11).
\]
A systematic trend on successively increasing $L_\mathrm{min}$ is not visible.

Regarding the domain-wall length, we again find only tiny scaling corrections, which
cannot be resolved for any $L > 20$. To avoid any risk from spurious remnant
corrections, we take $L_\mathrm{min} = 40$ for the uncorrected fit
$\langle \ell\rangle_J = A_\ell L^{d_\mathrm{f}}$ and arrive at
\[
  d_\mathrm{f} = 1.2732(5).
\]
which yields $Q=0.73$. This fit is shown together with the data in
Fig.~\ref{fig:ppbc_G}(c). Comparing the results for $\theta$ and $d_\mathrm{f}$
between the PFBC and PPBC cases we see that they are in perfect agreement with each
other, indicating that the results truly probe the asymptotic regime and acting as an
{\em ex post\/} verification of the correctness of the windowing technique for the
PPBC case.

\begin{figure}
  \begin{center}
    \includegraphics[width=0.95\columnwidth]{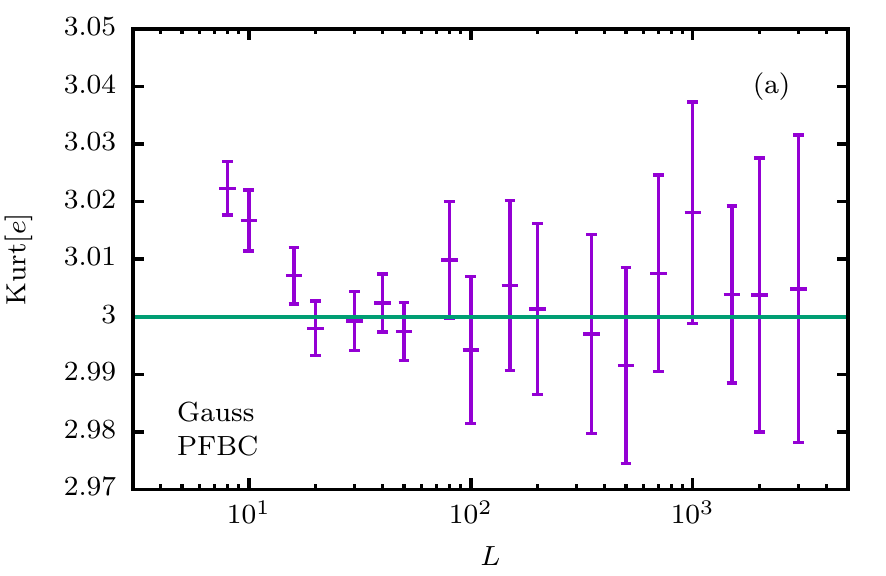}
    \includegraphics[width=0.95\columnwidth]{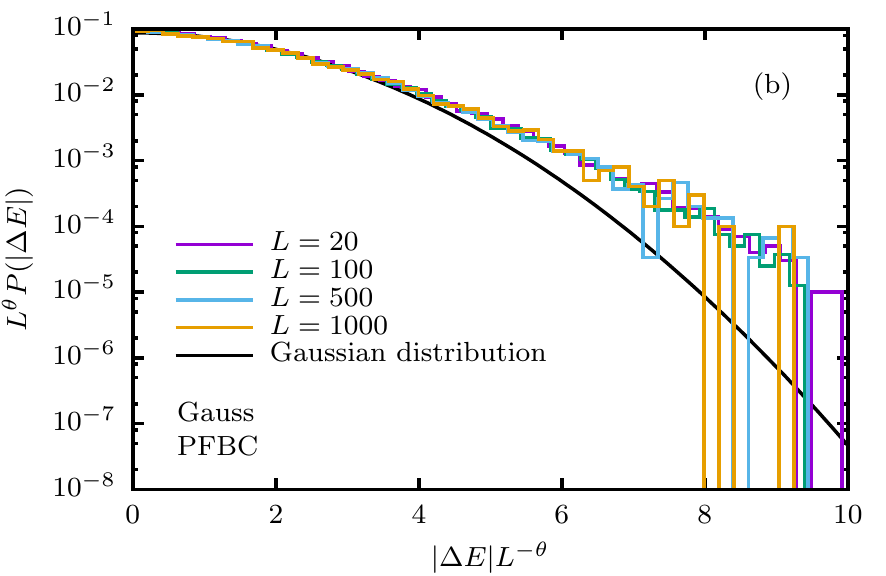}
    \includegraphics[width=0.95\columnwidth]{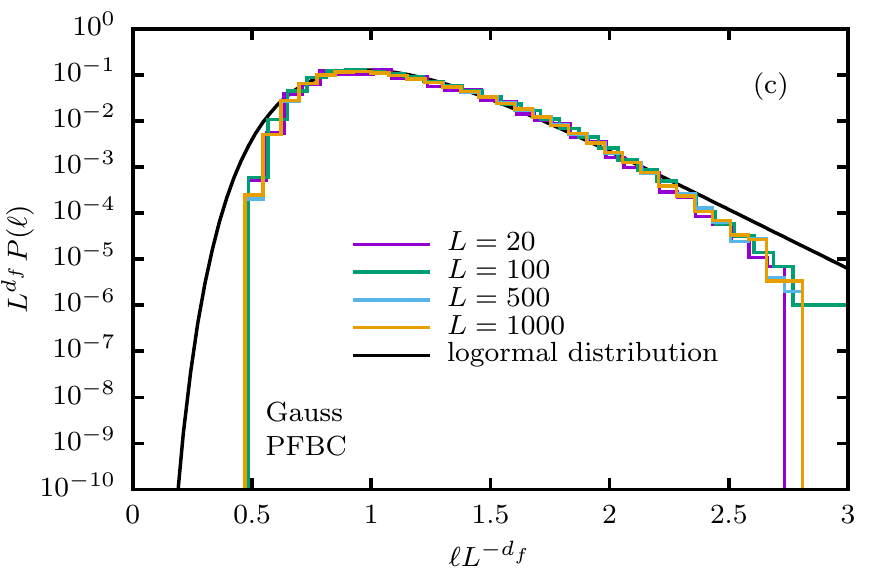}
    \caption{%
      (Color online) (a) Scaling of the kurtosis $\operatorname{Kurt}[e]$ of the
      distribution of ground-state energies per spin for the Gaussian model with PFBC
      as a function of system size. For $L\ge 20$ it is consistent with the value
      $\operatorname{Kurt}[e] = 3$ of a normal distribution. (b) Distribution of
      defect energies $|\Delta E|$ for the same model, rescaled with the expected
      asymptotic behavior $\propto L^{\theta}$ with $\theta = 0.2793$. The solid line
      shows a Gaussian distribution of the same mean and variance. (c) Distribution
      of domain-wall lengths $\ell$ for the PFBC Gaussian case, rescaled according to
      the limiting form $\propto L^{d_\mathrm{f}}$ with $d_\mathrm{f} =
      1.273\,19$.
      The solid line represents a lognormal distribution fitted to the empirical
      data.
    }
    \label{fig:distributions}
  \end{center}
\end {figure}

\subsection{Probability distributions}

When investigating the ground-state and defect energies as well as the domain-wall
lengths, besides looking at the average values reported above it is also instructive
to study the full distributions of these quantities over disorder samples. The width
$\langle (e-\langle e\rangle_J)^2\rangle_J$ of the distribution of ground-state
energies per spin shows power-law scaling according to $L^{\Theta_f}$, where we find
$\Theta_f = -0.9995(3)$ for PFBC and $\Theta_f = -1.002(1)$ for PPBC, consistent with
the theoretical expectation \cite{bouchaud:03} $\Theta_f = -d/2$. The latter follows
from a standard argument of decomposition of the system into effectively uncorrelated
subsystems, such that the total energy is a sum of independent contributions. As a
result, in the thermodynamic limit the distribution narrows to a delta peak,
consistent with the fact that the ground-state energy is self-averaging
\cite{dotsenko:17}. To investigate the shape of the distribution, we studied its
kurtosis defined in Eq.~\eqref{eq:kurtosis}. $\operatorname{Kurt}[e]$ is shown in
Fig.~\ref{fig:distributions}(a) for the PFBC case, where it is found to be consistent
with $3$ to within statistical errors for all lattice sizes $L\ge 20$, indicating
that the distribution of ground-state energies is in fact Gaussian
\cite{aspelmeier:03a}. This is in contrast to systems with long-range interactions
such as the Sherrington-Kirkpatrick model, where non-Gaussian distributions are found
\cite{bouchaud:03}.

For symmetric coupling distributions the histogram of defect energies for P and AP
boundaries is also symmetric and so has zero mean. It is expected that the standard
deviation $\sigma(E)$ has the same asymptotic scaling behavior as the modulus
$|\Delta E|$, and this is consistent with our observations. Considering the data for
$\sigma(\Delta E)$ for PFBC, we use a pure power-law fit with
$L\ge L_\mathrm{min} = 30$ to find $\theta = -0.2793(3)$ ($Q=0.55$). For PPBC, on the
other hand, the same analysis yields $\theta = -0.279(2)$ and $Q=0.81$ for the same
range. In Fig.~\ref{fig:distributions}(b) we show the defect energy distribution for
PFBC systems for a number of different lattice sizes, rescaled by the factor
$L^\theta$ with $\theta = -0.2793$ describing the decay in width. As the Gaussian
distribution with the same mean and width shows, the defect energy distribution is
clearly not normal, but instead has much heavier tails\footnote{We note that there
  might be a relation between the behavior of the defect-energy distribution at
  vanishing energies and the question of a multiplicity of states in spin glasses
  \cite{vaezi:17}.}. This is confirmed by an inspection of the distribution kurtosis,
$\operatorname{Kurt}[\Delta E]$, which is found to be consistent with
$\operatorname{Kurt}[\Delta E] = 4.70(2)$ for systems of size
$L\ge L_\mathrm{min} = 20$.

The standard deviation of the distribution of domain-wall lengths is found to have
the same scaling as the mean, i.e., it is asymptotically proportional to $L^{d_\mathrm{f}}$,
suggesting a complementary way of determining the fractal dimension. This approach
yields estimates of $d_\mathrm{f} = 1.2740(3)$ for PFBC ($L_\mathrm{min} = 40$, $Q=0.34$) and
$d_\mathrm{f} = 1.276(2)$ for PPBC ($L_\mathrm{min} = 50$, $Q=0.61$), respectively. The result
for PFBC is slightly high as compared to the result from the mean, but still
statistically consistent: the deviation is $2.6$ times the combined error bar, but
this does not take into account that the two error estimates are correlated and so
the combined fluctuation is likely higher than the naive estimate
\cite{weigel:09,weigel:10}. The two PPBC estimates are fully consistent. The
distribution of domain wall lengths is found to be clearly non-Gaussian, with a
kurtosis that is consistent with $\operatorname{Kurt}[\ell] = 3.656(4)$ for systems
of size $L\ge L_\mathrm{min} = 20$. It was suggested in Ref.~\onlinecite{melchert:07}
that the distribution might be in fact lognormal. Our data for the distribution of
$\ell$ for PFBC are shown in Fig.~\ref{fig:distributions}(c), together with a fit to
a lognormal distribution. As is apparent, it describes the data reasonably well close
to the mode, but there are significant deviations in the tails.

\section{Results for bimodal couplings\label{sec:bimodal}}

For bimodal couplings there is a huge ground-state degeneracy. As has been
demonstrated with numerical calculations \cite{blackman:91,saul:93} and also shown
rigorously \cite{avron:81}, this model even has a finite ground-state entropy,
indicating that the number of ground states grows exponentially with system size. It
turns out to be a challenge to fulfill the equilibrium requirement of ensuring that
all such states are sampled with equal probability.

\subsection{Uniform sampling of ground states}
\label{sec:gaussian-noise}

For the case of systems with ground-state degeneracies, the solution to the matching
problem described in Sec.~\ref{sec:window} is not unique. There are several, possibly
many solutions to the matching problem that have the same minimal weight. In
practice, the implementation of the matching algorithm used will return an arbitrary
solution out of this set, where the state chosen depends on the specific
implementation of the algorithm used (for instance on the order in which nodes and
edges are examined) and the state returned might or might not be reproducible between
runs\footnote{The energy of the state returned, on the other hand, is of course
  always the same.}. Clearly, this setup is not suitable for sampling such states
with a prescribed probability weight.

One way of solving this problem and ensuring uniform sampling of states might be to
break the degeneracy in a way such that each ground state is preferred the same
number of times by a chosen procedure. If one examines a pair of ground states, one
will find that they differ by the overturning of a set of disjoint, but singly
connected clusters of spins. As, by definition, this procedure does not change the
overall energy, this corresponds to a set of ``free'' spins \cite{khoshbakht:17}. The
degeneracy can be lifted by adding some small perturbation to the bonds, i.e.,
\begin{equation}
  \label{eq:gaussian_noise}
  J_{ij}(\kappa) = J_{ij} + \kappa \epsilon_{ij},
\end{equation}
with a continuous, symmetric distribution of the random variables $\epsilon_{ij}$, a
natural choice being the standard normal distribution,
$\epsilon_{ij} \sim {\cal N}(0,1)$. As the spectrum of states for the bimodal model
is gapped \cite{saul:93}, if $\kappa$ is chosen sufficiently small the ground state
of the system with couplings $J_{ij}(\kappa)$ will also be a ground state of the
system with $\kappa = 0$. Considering a cluster of free spins for a symmetric
distribution of $\epsilon_{ij}$, the sum of the noise terms $\epsilon_{ij}$ along the
bonds on the cluster boundary will have either sign with the same probability of
$1/2$. Hence one half of the realizations of $\epsilon_{ij}$ should lead to this
cluster being in one orientation and the other half to it being in the reversed
orientation, implying uniform sampling of degenerate ground states. A similar
approach was used in Refs.~\onlinecite{cieplak:90,gusman:08}. As we discuss elsewhere
\cite{khoshbakht:17}, however, clusters that touch each other are not independent and
hence the procedure leads to a strongly non-uniform distribution of sampled states.

Uniform sampling is achieved via a new technique based on a combination of
combinatorial optimization in the form of the MWPM algorithm and Markov chain Monte
Carlo \cite{khoshbakht:17}. We use MWPM to exactly determine the set of {\em rigid\/}
clusters in the ground-state manifold, i.e., the set of connected regions such that
the spins inside of them have the same relative orientation in all ground states. In
a second step, we then perform a parallel tempering simulation \cite{hukushima:96a}
with updates that are a combination of flipping individual rigid clusters and a
non-local cluster-update move \cite{houdayer:01}. Details of the procedure as well as
benchmarks will be presented elsewhere \cite{khoshbakht:17}.

\subsection{Ground-state and defect energies}

\begin{figure}
  \begin{center}
    \includegraphics[width=0.95\columnwidth]{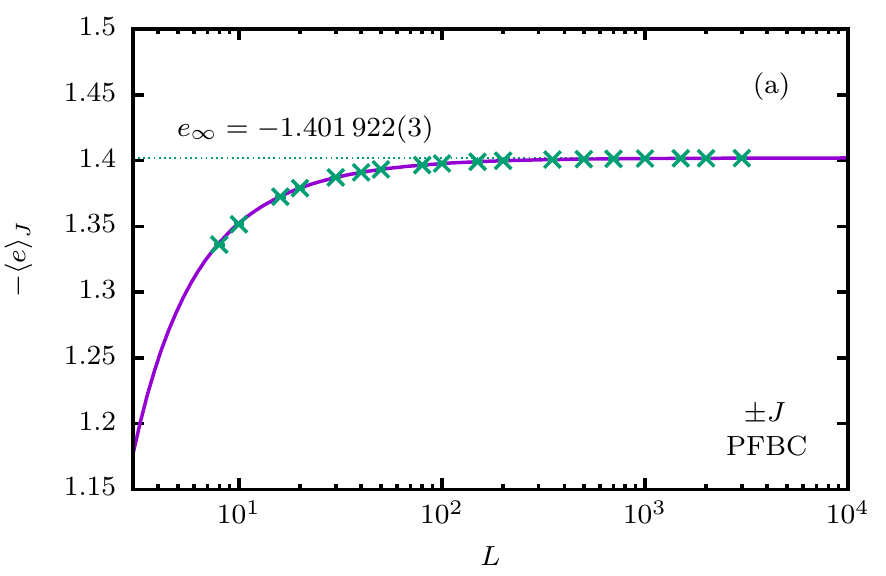}
    \includegraphics[width=0.95\columnwidth]{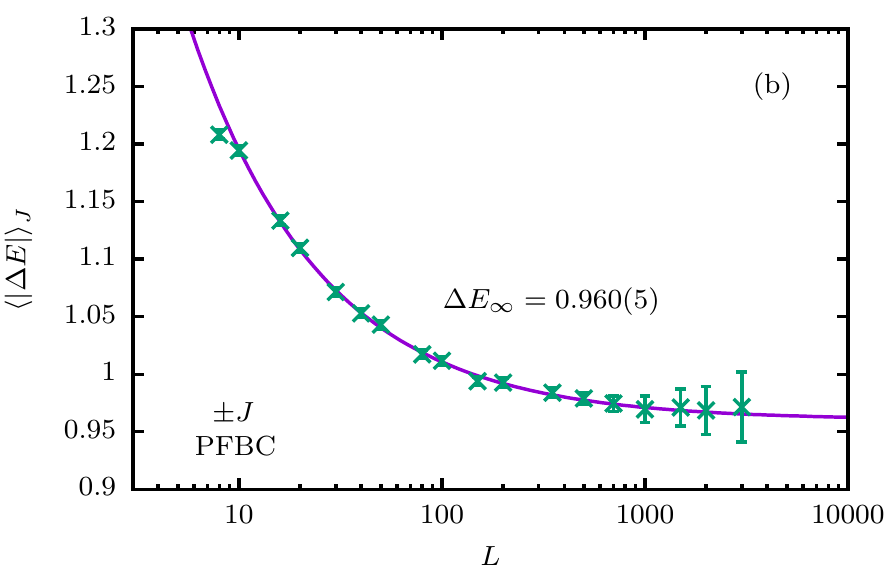}
    \includegraphics[width=0.95\columnwidth]{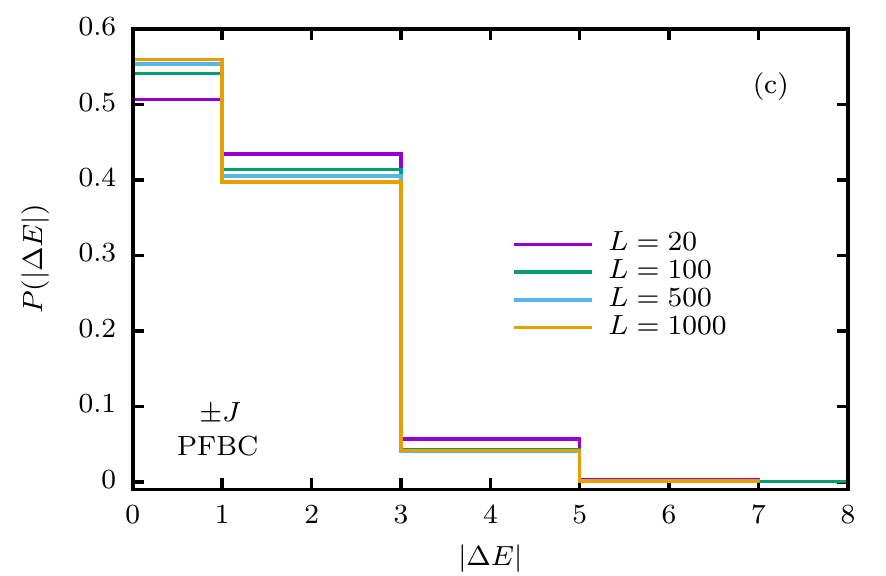}
    \caption{(Color online) (a) Average ground-state energies for bimodal couplings
      and PFBC boundaries, together with a fit of the functional form
      \eqref{eq:energy_pfbc} with $\theta = 0$ to the data for the range
      $L=20,\ldots,3000$. (b) Defect energies for systems with bimodal couplings and
      PFBC boundaries. Clearly, $\langle|\Delta E|\rangle_J$ converges to a non-zero
      value as $L\to\infty$, indicating that $\theta = 0$.  The line shows a fit of
      the form $\langle|\Delta E|\rangle_J =\Delta{E_\infty} + B_\theta L^{-\omega}$
      to the data with $L\ge L_\mathrm{min} = 10$ yielding
      $\Delta{E_\infty} = 0.960(5)$. (c) Probability distribution over disorder of
      the defect energies for the PFBC $\pm J$ model and different system sizes. For
      $L\to\infty$ the distribution approaches a limiting shape close to the $L=1000$
      case shown here.}
    \label{fig:pfbc_B}
  \end{center}
\end {figure}

For the ground-state energy the presence of degeneracies and sampling bias is not
relevant. We hence used the regular MWPM procedure to determine ground-state energies
for pairs of samples with periodic and antiperiodic boundaries and the resulting
defect energies. For these quantities we restricted our calculations to the case of
PFBC as this allows for treating larger system sizes, but studies of PPBC would also
be possible using the windowing technique. The range of system sizes and number of
realizations for each size are summarized in the fourth column of
Table~\ref{tab:samples}. The average ground-state energy per spin is shown in
Fig.~\ref{fig:pfbc_B}(a). Inspecting the general scaling ansatz
\eqref{eq:energy_pfbc} and taking into account that we expect $\theta = 0$ for this
model (see below), we should only have analytical corrections proportional to $1/L$
and $1/L^2$ up to $O(L^{-3})$, and indeed we find a good fit ($Q=0.18$) of this
functional form for the range $L \ge L_\mathrm{min} = 20$, yielding
\[
  e_\infty = -1.401\,922(3).
\]
This fit is shown together with the data in Fig.~\ref{fig:pfbc_B}(a). No drift of
$e_\infty$ is visible on further increasing $L_\mathrm{min}$.

The defect energies resulting from this procedure are shown in
Fig.~\ref{fig:pfbc_B}(b), indicating that for this model $\langle|\Delta E|\rangle_J$
converges to a finite value instead of decaying away to zero. This is consistent with
previous findings \cite{hartmann:01a,amoruso:03a}. If we assume a power-law decay as
prescribed by Eq.~\eqref{eq:defect-energy-corrected} and ignore the correction terms,
i.e., we use a pure power-law form $\langle|\Delta E|\rangle_J = A_\theta L^\theta$,
a good fit is achieved for $L\ge L_\mathrm{min} = 150$, resulting in
$\theta = -0.012(4)$, marginally compatible with $\theta = 0$. Additionally, the
modulus of $\theta$ systematically drops as $L_\mathrm{min}$ is increased. The defect
energy in this case hence does not decay to zero, but attains a non-zero value in the
thermodynamic limit. We therefore make the scaling ansatz
\begin{equation}
  \langle|\Delta E|\rangle_J = \Delta{E_\infty} + B_\theta L^{-\omega}.
  \label{eq:defect-energy-bimodal}
\end{equation}
We find an excellent fit with $Q=0.99$ already for $L_\mathrm{min} = 10$, resulting
in
\[
 \Delta{E_\infty} = 0.960(5) 
\]
and $\omega = 0.67(4)$. An alternative fit form including analytic corrections
proportional to $1/L$ and $1/L^2$ but omitting the $ L^{-\omega}$ term is found to be
of significantly lower quality.

Studying the distributions of both ground-state and defect energies, we again find a
Gaussian shape for the ground-state energies, the kurtosis being compatible with that
of a normal distribution for all system sizes studied. The standard deviation of the
defect energy shows analogous behavior to $\langle|\Delta E|\rangle_J$, settling down
at a finite value as $L\to\infty$. A fit of the form \eqref{eq:defect-energy-bimodal}
yields an asymptotic $\sigma_\infty(\Delta E) = 1.1564(4)$ ($L_\mathrm{min} = 16$,
$Q=0.41$). The disorder distribution of defect energies is shown in
Fig.~\ref{fig:pfbc_B}(c), illustrating that it approaches a limiting shape as
$L\to\infty$ in which about 57\% of domain walls have zero energy, 38\% have
$\Delta E = 2$, 4\% have $\Delta E = 4$, and higher defect energies occur in less
than 1\% of the cases.

\subsection{Domain walls}

\begin{figure}
  \begin{center}
    \includegraphics[width=0.95\columnwidth]{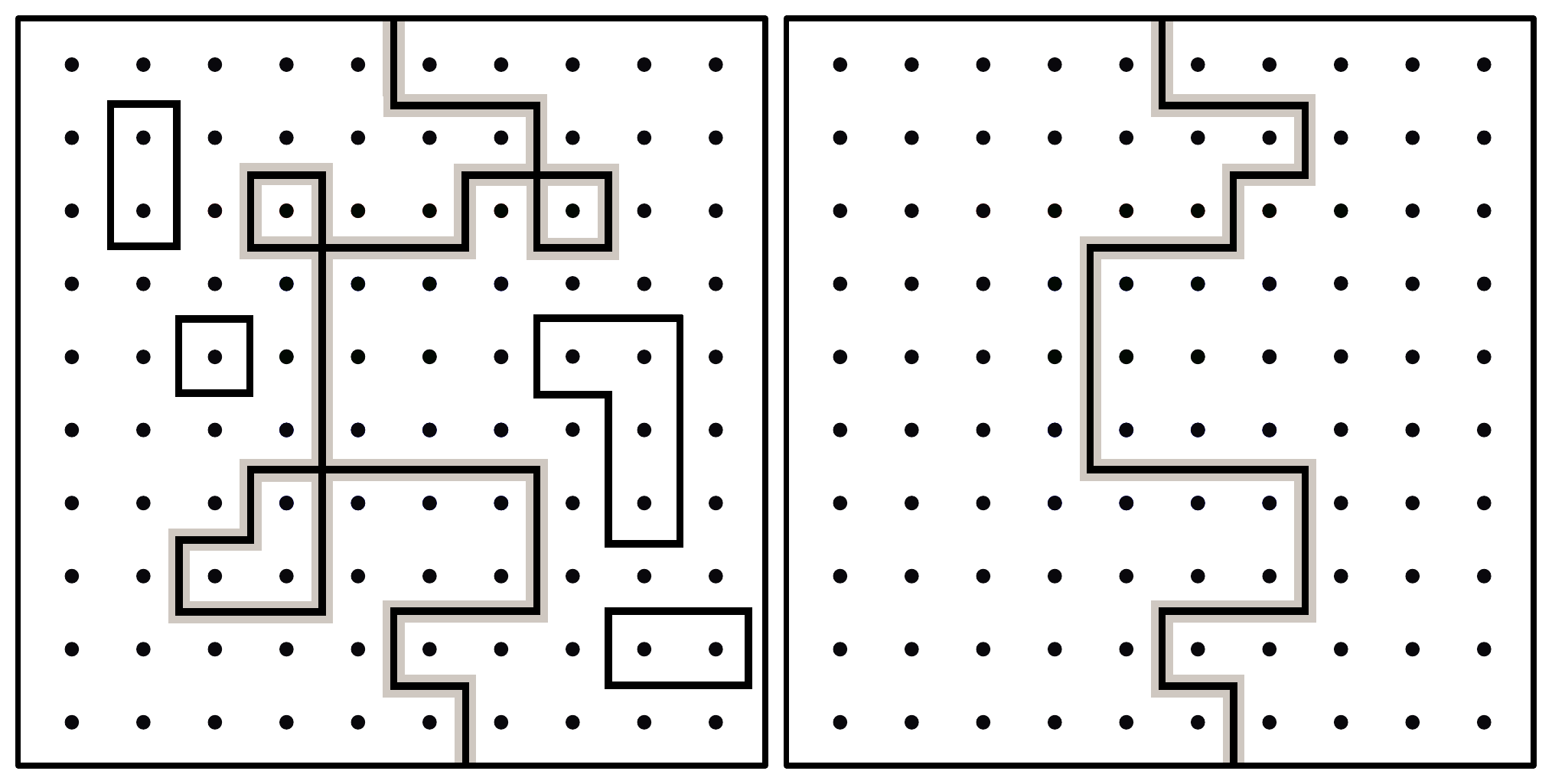}
    \caption{%
      Left: Schematic representation of the set of dual bonds satisfying the
      condition \eqref{eq:dw-def} for the case of bimodal couplings. Besides the
      domain wall it contains isolated loops enclosing free clusters of spins as well
      as bubbles of free spins attached to the domain wall. Removing the isolated
      free loops one arrives at the set ${\cal D}_\mathrm{long}$, which we denote as
      the ``long'' domain wall. Right: After the additional removal of bubbles one
      arrives at the set ${\cal D}_\mathrm{short}$ of dual bonds comprising the
      ``short'' domain wall of the configuration.%
    }
    \label{fig:domain_wall}
  \end{center}
\end {figure}

The presence of free clusters of spins in the manifold of degenerate ground states
complicates the identification of domain walls for the bimodal model
\cite{gusman:08}. A possible difference in configuration between the ground state for
a disorder configuration with P boundaries and a ground state for the same
realization with AP boundary conditions is schematically depicted in the left panel
of Fig.~\ref{fig:domain_wall}. We see that in this case the set of domain-wall bonds
satisfying condition \eqref{eq:dw-def}, i.e., different relative orientations of
spins at both ends for the P and AP configurations, does not only contain the actual
domain wall but also a set of closed loops detached from the wall. These correspond
to free clusters that can be overturned at zero energy cost and so happen to be in
one orientation in the P ground state, but in the opposite orientation in the AP
configuration. Conceptually, these bonds do not belong to the domain wall. We remove
them by only counting the system spanning part of the set ${\cal D}$. We refer to the
corresponding set, denoted as ${\cal D}_\mathrm{long}$, as the ``long'' domain wall
and its length as $\ell_\mathrm{long} = |{\cal D}_\mathrm{long}|$. Additionally,
however, it is possible for such free clusters to be attached to the domain wall as
is also depicted in the example of Fig.~\ref{fig:domain_wall}. Such ``bubbles''
attached to corners of the wall are somewhat arbitrary additions and removing them by
only considering the shortest path in the set ${\cal D}$ connecting opposite ends of
the system defines the reduced set ${\cal D}_\mathrm{short}$ with
$\ell_\mathrm{short} = |{\cal D}_\mathrm{short}|$. Clearly we have that
${\cal D}_\mathrm{short} \subseteq {\cal D}_\mathrm{long } \subseteq {\cal D}$. Note
that even after these removals the set ${\cal D}_\mathrm{short}$ is not unique for a
given bond configuration, and the additional degeneracy is connected to zero-energy
loops that share (at least) one bond with the domain wall (instead of only sharing a
corner) and hence can be interpreted as diversions of the wall. In order to probe the
equilibrium properties, we must sample from such walls with equal probability.

% \begin{table}[tb!]
%   \caption{The number of disorder realizations for different boundary conditions and
%     system sizes considered for the uniform sampling algorithm of
%     Ref.~\onlinecite{khoshbakht:17}.}
%   \begin{ruledtabular}
%     \begin{tabular}{c c c c}
%      & $L$   &   PFBC Bimodal &  \\
%       \hline                   
%      & 10    &   $1\times10^3$ & \\
%      & 16    &   $1\times10^3$ & \\
%      & 20    &   $1\times10^3$ & \\
%      & 24    &   $1\times10^3$ & \\
%      & 28    &   $1\times10^3$ & \\
%      & 32    &   $1\times10^3$ & \\
%      & 48    &   $1\times10^3$ & \\
%      & 64    &   $1\times10^3$ & \\
%      & 80    &   $1\times10^3$ & \\
%      & 100   &   $1\times10^3$ & \\
%      & 128   &   $1\times10^3$ & \\ 
% %      156   &   $2\times10^3$  \\ 
%     \end{tabular}
%   \end{ruledtabular}
%   \label{tab:uniform_samples}
% \end{table}

\begin{figure}[tb!]
  \begin{center}
    \includegraphics[width=0.95\columnwidth]{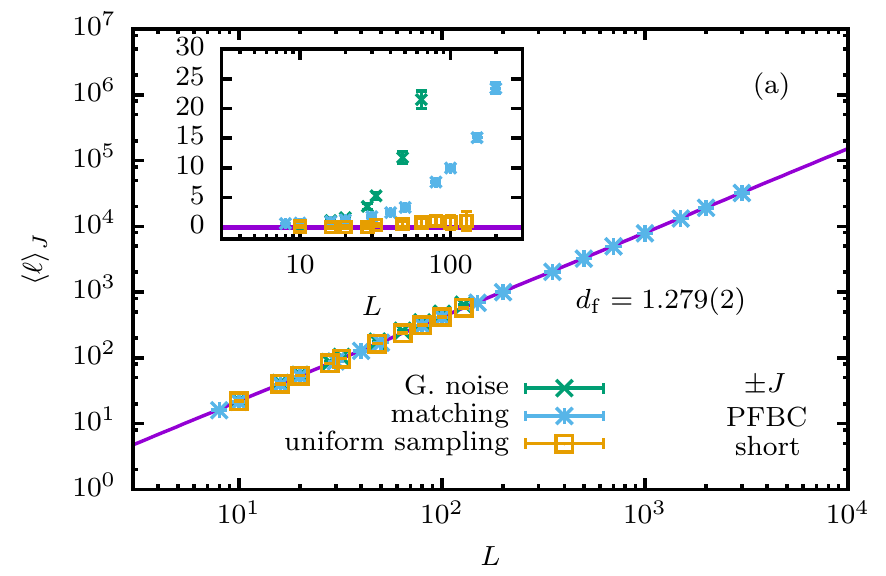}
    \includegraphics[width=0.95\columnwidth]{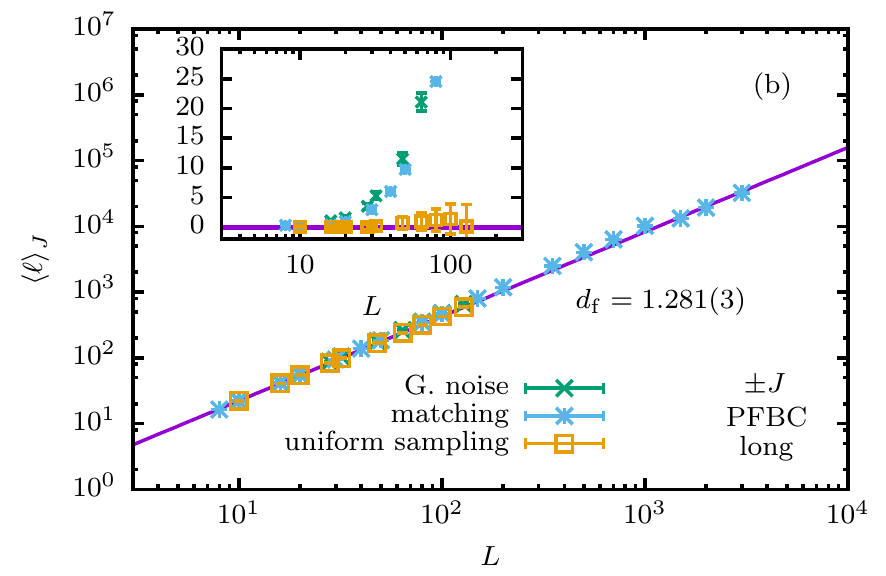}
    \includegraphics[width=0.95\columnwidth]{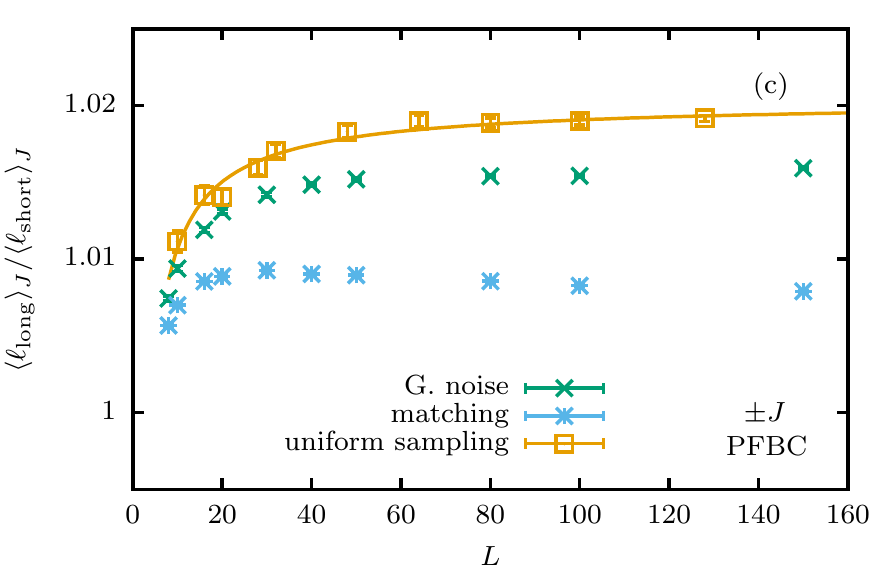}
    \caption{(Color online)%
      (a) Average length $\langle\ell_\mathrm{short}\rangle_J$ of the short domain
      wall for the bimodal model as a function of linear system size $L$ for the
      three different algorithms employed. The inset shows the deviation of each data
      set from the fit of the power law
      $\langle\ell\rangle_J = A_\ell L^{d_\mathrm{f}}$ to the uniform sampling data
      for $L\ge L_\mathrm{min} = 16$, which results in $d_\mathrm{f} = 1.279(2)$
      ($Q=0.33$). (b) Average length $\langle\ell_\mathrm{long}\rangle_J$ of the long
      domain walls for the different algorithms.  The inset shows the deviation of
      each data set from the fit of a pure power law to the uniform data, yielding
      $d_\mathrm{f} = 1.281(3)$ for $L_\mathrm{min} = 16$ ($Q=0.97$). (c) Ratio of
      the average lengths of long and short domain walls as estimated from the
      different algorithms. In all cases, the ratio approaches a constant, in line
      with the identical estimates of fractal dimension for $\ell_\mathrm{short}$ and
      $\ell_\mathrm{long}$.
    }
    \label{fig:uniform-sampling}
  \end{center}
\end {figure}

Regarding the sampling of domain-wall lengths for the bimodal model we have produced
data from three different algorithms:
\begin{enumerate}
\item Our implementation of the MWPM algorithm calculates a ground-state for each
  sample with both P and AP boundary conditions, and comparing these we can determine
  the lengths $\ell_\mathrm{short}$ and $\ell_\mathrm{long}$ of the related domain
  walls. It is clear that this does not correspond to a fair sampling of ground
  states, but the nature of the bias depends on internal details of the MWPM
  implementation \cite{kolmogorov:09} and is not clear on a physical level. This
  technique allows to treat large system sizes and we applied it to the data set of
  sizes $8\le L\le 3000$ described in the third column of Table \ref{tab:samples}. In
  the following, we denote this as the ``matching'' algorithm.
\item The Gaussian noise technique described in Sec.~\ref{sec:gaussian-noise} is
  designed to break the degeneracy in a systematic way. For each realization it only
  requires an additional run of the MWPM algorithm per boundary condition, and we
  hence applied it to the same set of samples with $8\le L\le 3000$. As discussed in
  Sec.~\ref{sec:gaussian-noise} it also does not provide uniform samples,
  however. This technique is referred to as ``Gaussian noise'' in the following.
\item The new algorithm based on a cluster decomposition and parallel tempering
  outlined in Ref.~\onlinecite{khoshbakht:17} provides uniform samples, but it is
  much more demanding computationally, such that only smaller system sizes can be
  treated reliably. We have studied systems of edge lengths $L=10$, $16$, $20$, $24$,
  $28$, $32$, $48$, $64$, $80$, $100$, and $128$ for this method, using $1000$
  samples per size and producing ten independent ground-state configurations per
  sample.
  % Here, additional to the data set with PFBC we also studied samples with PPBC
  % using the windowing technique.
  Data from this algorithm are labeled ``uniform sampling''.
\end{enumerate}

Figure \ref{fig:uniform-sampling}(a) shows the three data sets for the scaling of the
lengths of short domain walls. On the scale of the domain-wall lengths
themselves, all data appear to fall on top of each other, but a closer inspection
reveals that this is in fact not the case. The data from  uniform sampling show very
clean scaling behavior and a pure power law $\langle\ell\rangle_J = A_\ell
L^{d_\mathrm{f}}$ describes the data for $L\ge L_\mathrm{min} = 16$ well. No drift of
the exponent value is observed on omitting further values on  the small-$L$ side. The
fractal dimension is estimated from this fit as
\[
  d_\mathrm{f} = 1.279(2)
\]
with $Q=0.33$. As the inset of Fig.~\ref{fig:uniform-sampling}(a) shows, there are
statistically significant deviations of the data from the other two sampling
techniques from this result. The samples generated by the Gaussian noise technique
show clean scaling as well, but with a significantly larger exponent
$d_\mathrm{f} = 1.323(3)$ ($L_\mathrm{min} = 16$, $Q=0.86$). The data from the
matching approach alone, on the other hand, show somewhat inconsistent behavior for
successive system sizes, and they are compatible with a pure power law only for
$L \ge L_\mathrm{min} = 80$, yielding $d_\mathrm{f} = 1.2802(5)$ ($Q=0.18$). This
slightly unsteady statistical behavior is probably connected to the fact that the
matching technique does not use a stochastic sampling technique, and due to internal
design decisions the behavior of the algorithm might change discontinuously at
certain system sizes. Somewhat surprisingly, however, the results for the pure
matching technique are closer to the correct result represented by uniform sampling
than the samples produced by Gaussian noise, see also the inset of
Fig.~\ref{fig:uniform-sampling}(a).

We move on to considering the results for the long domain walls. The data are
summarized in Fig.~\ref{fig:uniform-sampling}(b). While for each data set, the values
of $\langle \ell_\mathrm{long}\rangle_J$ are somewhat larger than those of
$\langle \ell_\mathrm{short}\rangle_J$ the relative behavior of the three data sets
for the long domain walls is very similar to that found for the short walls. From the
uniform sampling data, a pure power-law fit for $L_\mathrm{min}=16$ yields
$d_\mathrm{f} = 1.281(3)$ ($Q=0.97$) which is statistically consistent with the
result from the short domain walls. For comparison, matching and Gaussian noise yield
$d_\mathrm{f} = 1.2797(5)$ and $d_\mathrm{f} = 1.325(3)$, respectively, for the same
ranges that were used for the short walls. It hence appears that for the scaling of
domain-wall length with system size, there is no difference between the short and
long definitions of domain walls. This impression is corroborated by the data shown
in Fig.~\ref{fig:uniform-sampling}(c) of the ratios of long and short lengths of
domain walls, averaged over disorder, for the three different techniques. It is clear
that this ratio settles down to a finite value as $L\to\infty$, and a fit of the
function form
$\langle\ell_\mathrm{long}/\ell_\mathrm{short}\rangle_J = \kappa + A_\kappa
L^{-\omega}$
to the uniform sampling data yields $\kappa = 1.021(6)$ and $\omega = 0.85(16)$ with
$Q=0.18$ ($L_\mathrm{min}=10$).

\begin{figure}
  \begin{center}
    \includegraphics[width=0.95\columnwidth]{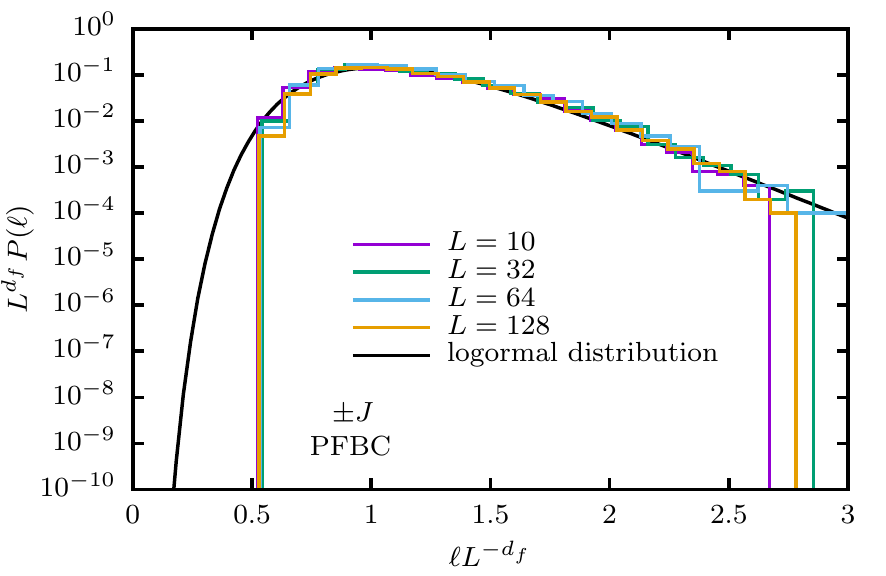}
    \caption{(Color online)%
      Distribution of the lengths $\ell_\mathrm{long}$ of the long domain walls for
      $\pm J$ couplings and PFBC boundaries as resulting from the uniform sampling
      approach. The re-scaling of the axes is with respect to the fractal dimension
      $d_\mathrm{f} = 1.281(3)$ estimated from the data in
      Fig.~\ref{fig:uniform-sampling}.
    }
    \label{fig:bimodal_DW_distribution}
  \end{center}
\end {figure}

It is worthwhile to compare these estimates of the fractal dimension to those found
previously: Melchert and Hartmann \cite{melchert:07} used combinatorial optimization
methods to find minimal and maximal domain walls in the manifold of degenerate
ground-state pairs, yielding lower and upper bounds for $d_\mathrm{f}$, namely
$1.095(2) \le d_\mathrm{f} \le 1.395(3)$. Our estimates are clearly compatible with
these, and it is interesting to note that the actual value is much closer to the
upper than to the lower limit which corresponds to almost flat walls.  Risau-Gusman
and Rom\'a \cite{gusman:08} estimate $d_\mathrm{f} = 1.323(3)$ using non-uniform
sampling resulting from employing the bare MWPM algorithm; this is compatible with
our ``matching'' results, but too large compared to the unbiased estimate from
uniform sampling. Studying domain walls in a hexagonal lattice, Weigel and Johnston
\cite{weigel:06a} find $d_\mathrm{f} = 1.283(11)$, but again not using unbiased
sampling. Analyzing the behavior of the ground-state entropy, Fisch \cite{fisch:08}
estimates $d_\mathrm{f} = 1.22(1)$ which is strongly incompatible with our results,
which could be a sign of the relation $d_\mathrm{f} = 2\theta_S$ on which Fisch's
estimate is based, where $\theta_S$ is the scaling exponent of the ground-state
entropy, not being valid in two dimensions.

We finally tend to the distribution of domain-wall lengths for this case. As is
illustrated in Fig.~\ref{fig:bimodal_DW_distribution} for the long domain walls,
these follow the scaling form
$P(\ell) = L^{d_\mathrm{f}} \hat{P}(\ell L^{-d_\mathrm{f}})$ already observed for the
case with Gaussian couplings, cf.\ Fig.~\ref{fig:distributions}(c), where now
$d_\mathrm{f} = 1.281(3)$. The fit to a log-normal distribution also shown in
Fig.~\ref{fig:bimodal_DW_distribution} works quite well over the full range of the
distribution, in contrast to the case of Gaussian couplings, where deviations could
be seen in the right tail, cf.~Fig.~\ref{fig:distributions}(c). Very similar results
are obtained for the distribution of short domain walls also (not shown).

\section{Conclusions\label{sec:conclusions}}

We used an exact algorithm based on minimum-weight perfect matching to calculate
ground states for the square-lattice Ising spin glass with Gaussian and bimodal
couplings and lattice sizes of up to $10\,000\times 10\,000$ ($10^8$) spins,
employing periodic boundary conditions in one direction and free boundaries in the
other. For systems with full periodic boundaries, we developed a quasi-exact
algorithm that can find true ground-states with arbitrarily high probability and a
computational effort that is a constant time larger than for the planar graphs, and
we used it to study systems of up to $3\,000\times 3\,000$ spins. Our estimates of
the ground-state energies $e_\infty = -1.314\,787\,6(7)$ (Gaussian model) and
$e_\infty=-1.401\,922(3)$ (bimodal model) are compatible with, but up to 100 times
more precise than the estimates in the careful study of
Ref.~\onlinecite{hartmann:04a} using exact ground-state methods and the recent work
Ref.~\onlinecite{perez-morelo:13} using Monte Carlo. For Gaussian couplings, we also
determined the spin-stiffness exponent and the fractal dimension of domain walls with
unprecedented precision, yielding $\theta = -0.2793(3)$ and
$d_\mathrm{f} = 1.273\,19(9)$. These estimates are one to two orders of magnitude
more precise than previous results, see the data collected in Table
\ref{theta_articles}. We note that this value is also consistent with the most recent
estimate of $1/\nu = -\theta = 0.283(6)$ in Ref.~\onlinecite{fernandez:16}, but the
zero-temperature result has 10-fold increased precision. For bimodal couplings, we
find $\theta = 0$, in agreement with previous studies. Due to the large degeneracy of
the ground state for bimodal couplings, methods based on matching do not allow to
sample states with the proper statistical weight, and as a result unbiased estimates
of the domain-wall fractal dimension have not been possible previously. Using a newly
developed algorithm \cite{khoshbakht:17} allowed us to sample exact ground states for
this case uniformly, here up to system size $L=128$. The resulting estimates of the
fractal dimension, $d_\mathrm{f} = 1.279(2)$ and $d_\mathrm{f} = 1.281(3)$ for
``short'' and ``long'' domain walls, respectively, are marginally consistent with
$d_\mathrm{f}$ for the Gaussian couplings, the deviation being 3 and 4 standard
deviations, respectively.

In 2006, Amoruso {\em et al.\/} \cite{amoruso:06a} used results from stochastic
Loewner evolution (SLE) to conjecture that the 2D spin glass with Gaussian couplings
is described by a non-unitary conformal field theory with central charge $c < -1$,
related to the SLE parameter $\kappa$ as \cite{henkel:12}
$c = (6-\kappa)(3\kappa-8)/\kappa$, and they numerically determined a value
$\kappa \approx 2.1$. Further, it was assumed that the scaling dimension
$x_t = d-y_t = d-1/\nu = d+\theta = 2+\theta$ of the energy operator should be
represented in the corresponding Kac table \cite{henkel:book}, and a numerically
close value was found in tentatively identifying
$x_t = 2\Delta_{1,2} = (6-\kappa)/\kappa$. Together with the relation
$d_\mathrm{f} = 1+\kappa/8$ for the fractal dimension, this yields the equation
\cite{amoruso:06a}
\begin{equation}
  \label{eq:sle-relation}
  d_\mathrm{f} = 1+\frac{3}{4(3+\theta)}.
\end{equation}
We note that additional to the assumption of a CFT representation, the identification
of conformal weights with items in the Kac table is only supposed to work for
rational values of $\kappa$, which does not appear to be the case
here. Eq.~\eqref{eq:sle-relation} was found to be consistent with previous estimates
of $\theta$ and $d_\mathrm{f}$. \cite{amoruso:06a} Our most accurate results are for
PFBC boundaries. The corresponding estimate $d_\mathrm{f} = 1.273\,19(9)$ would imply
via Eq.~\eqref{eq:sle-relation} that $\theta = -0.2546(9)$ which does not seem
consistent with the estimate $\theta = -0.2793(3)$ from the defect energies. More
systematically, if \eqref{eq:sle-relation} is to hold, the difference
\[
  d_\mathrm{f} -1 -\frac{3}{4(3+\theta)} = -0.00247(9)
\]
must be consistent with zero. Here, we used the estimates for $d_\mathrm{f}$ and
$\theta$ from PFBC and standard error propagation \cite{brandt:book}. The difference
from zero corresponds to about 27 standard deviations, so based on the usual
confidence limits one would need to reject the hypothesis that our data are
consistent with \eqref{eq:sle-relation}. This neglects the fact, however, that our
estimates for $d_\mathrm{f}$ and $\theta$ are correlated as they are derived from the
same set of disorder realizations \cite{weigel:10}. To correct for this effect, we
divided the disorder samples for PFBC such that one half is used to estimate
$\theta=-0.2795(3)$ ($Q=0.47$) and the other half is used to estimate
$d_\mathrm{f} = 1.273\,22(12)$ ($Q=0.32$) using the same fit functions and ranges as
for the full data set. With these estimates, we find
\[
  d_\mathrm{f} -1 -\frac{3}{4(3+\theta)} = -0.00246(12),
\]
where the deviation from zero is still about 20 standard deviations, corresponding to
the expected reduction by halving the statistics, so the correlation effect appears
to be weak. As an alternative analysis, we also attempted to perform a simultaneous
fit of power laws to the scaling of $|\Delta E|$ and $\ell$ while enforcing the
relation \eqref{eq:sle-relation} between the scaling exponents. Independent of
whether we use the full or the split data set, a fit quality $Q> 0.01$ is only
achieved for $L_\mathrm{min} \ge 1000$, which is way above the range of lattice sizes
where scaling corrections are visible above the statistical errors (recall that both
the defect energies and domain-wall lengths are fully consistent statistically with
pure power-laws for $L > L_\mathrm{min} = 40$). The conclusions from considering the
independent data set for PPBC are similar, with the deviation from
Eq.~\eqref{eq:sle-relation} being $-0.00231(47)$, corresponding to 5 standard
deviations. The values for the deviations for PFBC and PPBC are statistically
consistent, the appearance of better consistency for PPBC is due to the smaller
statistics there. While it is always difficult to reject or confirm an exact (but
non-rigorous) relation based on numerics, it appears safe to say that our data do not
appear to be consistent with Eq.~\eqref{eq:sle-relation}\footnote{While we tried to
  take careful account of scaling corrections by including additional terms in the
  fit functions and/or monitoring the dependence of the results on the choice of
  $L_\mathrm{min}$, it is not possible to completely exclude the possibility of
  spurious systematic corrections leading to the observed deviations from
  Eq.~\eqref{eq:sle-relation}.}. It is worthwhile to note that, on the other hand,
our values for $\theta$ and $d_\mathrm{f}$ are fully consistent with previous
estimates, cf.\ the data compiled in Table \ref{theta_articles}, and it is only due
to the increased accuracy resulting from the bigger systems and larger numbers of
disorder samples considered here that the inconsistency with
Eq.~\eqref{eq:sle-relation} arises.

Our results for the fractal dimension of the bimodal model are marginally consistent
with those for the Gaussian model, and it remains an interesting question for further
studies whether universality between the two models holds in this respect.

\begin{acknowledgments}
  We are grateful to Frank Beyer, Giorgio Parisi, Michael Moore, and Jacob Stevenson
  for useful discussions. MW acknowledges extensive discussions with Zohar Nussinov,
  Gerardo Ortiz and Mohammad-Sadegh Vaezi on the physics of the spin-glass phase. We
  acknowledge funding from the DFG in the Emmy Noether Programme (WE4425/1-1) and
  funding from the European Commission through the IRSES network DIONICOS
  (PIRSES-GA-2013-612707).
\end{acknowledgments}

%\bibliography{citeulike_nourl_noissn}
\bibliography{PRE}

\end{document}